\documentclass[11pt,a4paper,preprintnumbers,nofootinbib]{revtex4}

\pdfoutput=1

\usepackage[T1]{fontenc}
\usepackage[utf8]{inputenc}
\usepackage{amsmath}
\usepackage{amssymb}
\usepackage{hyperref}
\usepackage[final]{graphicx}

\usepackage{ulem}

\begin{document}

\title{Rare radiative charm decays within the standard model and beyond}
\author{Stefan de Boer}
\email{stefan.deboer@tu-dortmund.de}
\author{Gudrun Hiller}
\email{ghiller@physik.uni-dortmund.de}
\affiliation{Fakultät für Physik, TU Dortmund, Otto-Hahn-Str.4, D-44221 Dortmund, Germany}
\begin{abstract} 
We present standard model (SM)  estimates for  exclusive $c \to u \gamma$ processes 
in heavy quark and hybrid  frameworks. 
Measured branching ratios ${\cal{B}}(D^0 \to (\phi, \bar K^{*0}) \gamma)$  are   at or somewhat exceeding the upper range of the SM 
and suggest slow convergence of the $1/m_D, \alpha_s$-expansion.
Model-independent constraints on $|\Delta C|=|\Delta U|=1$ dipole operators from ${\cal{B}}(D^0 \to  \rho^0  \gamma)$ data are obtained.
Predictions and implications for   leptoquark models  are worked out. While branching ratios are SM-like CP asymmetries
 $\lesssim 10 \%$ can be induced. In SUSY deviations from the SM  can be even larger with CP asymmetries of  $O(0.1)$.
If  $\Lambda_c$-baryons are produced polarized, such as at the $Z$-pole,  an angular asymmetry in $\Lambda_c \to p \gamma$ decays  can
be studied that is sensitive to chirality-flipped contributions.
\end{abstract}

\preprint{DO-TH 16/20, QFET-2016-18}

\maketitle

\section{Introduction}

A multitude of radiative charm decays is accessible at current and future high luminosity flavor facilities  \cite{Hewett:2004tv,Aushev:2010bq}.
In anticipation of the new  data we revisit  opportunities to test the  standard model (SM) with $c \to u \gamma$ transitions 
 \cite{Burdman:1995te,Khodjamirian:1995uc,Greub:1996wn,Fajfer:1997bh,Fajfer:1998dv,Fajfer:1999dq,Fajfer:2000zx,Isidori:2012yx}, complementing studies with  dileptons, {\it e.g., }\cite{Paul:2011ar,Fajfer:2001sa,deBoer:2015boa}.
To estimate the beyond the standard model (BSM) reach we detail and evaluate  exclusive $D_{(s)} \to V \gamma$ decay amplitudes, where $V$ is a  light vector meson. We employ   two frameworks, one based on the heavy quark expansion and QCD, adopting  expressions   from $b$-physics \cite{Bosch:2001gv},  and  a  hybrid phenomenological one, updating  \cite{Fajfer:1997bh,Fajfer:1998dv}. The latter combines chiral perturbation and  heavy quark effective theory and vector meson dominance (VMD).
Both frameworks  have considerable  systematic uncertainties, leaving
  individual charm branching ratios without clear-cut interpretation unless the deviation from the SM becomes somewhat obvious.
 On the other hand, considering several observables,  correlations can shed light on  hadronic parameters or on the electroweak model \cite{Fajfer:2000zx}.
The interpretation of  asymmetries is much easier, as (approximate) symmetries of the SM make them negligible 
compared to the  experimental precision for a while. 
In particular, we discuss implications of the recent measurements by  Belle  \cite{Abdesselam:2016yvr} 
\begin{align} \nonumber 
 {\cal{B}}(D^0 \to \rho^0 \gamma) & = (1.77\pm0.30 \pm 0.07)\cdot10^{-5}   \, , \\
\label{eq:ACP}
 A_{CP}(D^0 \to \rho^0 \gamma) & 
 = 0.056\pm 0.152\pm 0.006 \, ,
\end{align}
where the CP asymmetry $A_{CP}$ is defined as\footnote{The CP asymmetry of $D^0 \to \rho^0 \gamma$  is mostly direct, analogous to the time-integrated CP asymmetry in $D^0\to K^+K^-$ \cite{Aaij:2016cfh}.
We thank Alan Schwartz for providing us with this information.
In this work, we refer to $A_{CP}$ as the direct CP asymmetry, neglecting the small indirect contribution.}
\begin{align}
 A_{CP}(D \to V \gamma) & =\frac{\Gamma(D \to V \gamma)-\Gamma(\bar D \to\bar V \gamma)}{\Gamma(D\to V \gamma)+\Gamma(\bar D \to\bar V \gamma)} \, .
 \end{align}
We compare  data (\ref{eq:ACP})  to the SM predictions  and derive model-independent constraints on BSM couplings.
We further discuss  two specific BSM scenarios, leptoquark models and the minimal supersymmetric standard model with flavor mixing (SUSY).
For the former we point out that large  logarithms from the leading 1-loop diagrams with leptons and leptoquarks require resummation. 
The outcome is numerically of relevance for the interpretation of  radiative charm decays.

We further obtain analytical expressions for the contributions from the QCD-penguin operators to the effective dipole coefficient at 2-loop QCD.
This extends the description of radiative and semileptonic  $|\Delta C|=|\Delta U|=1$ processes at this order  \cite{Greub:1996wn,deBoer:2015boa,deBoer:2016dcg}.

While one expects the heavy quark and $\alpha_s$-expansion to perform worse  than in $b$-physics 
an actual quantitative evaluation of the individual contributions in radiative charm decays has not been done to date.
Our motivation is to fill this gap and detail the 
expansion's performance when compared to the hybrid model, and to data. In view of the importance of charm for probing flavor in and beyond the SM 
seeking  after opportunities for any, possibly data-driven improvement of the theory-description is worthwhile.

The organization of this paper is as follows:
In section \ref{sec:SM}  we  calculate weak annihilation and hard scattering contributions  to $D\to V\gamma$  decay amplitudes. In section \ref{sec:pheno} we  present SM predictions for branching ratios  and CP asymmetries in this approach and  in the hybrid  model.
We present model-independent constraints on BSM physics and look into  leptoquark models and  SUSY within the  mass insertion approximation in section  \ref{sec:BSM}.
Section \ref{sec:pol} is on $\Lambda_c\to p\gamma$ decays and the testability of a polarized $\Lambda_c$-induced angular asymmetry at future colliders \cite{Zhao:2002zk,dEnterria:2016fpc}.
In section \ref{sec:con} we summarize.
In appendix \ref{app:parameters} and \ref{app:DV_form_factors} we give the numerical input and $D \to V$ form factors used in our analysis. 
Amplitudes in the hybrid model are provided in appendix \ref{app:resonant_DVgamma_amplitudes}. Details on the 2-loop contribution from QCD-penguin operators are given in
appendix \ref{app:C7eff}.

\section{\texorpdfstring{$D\to V\gamma$}{DtoVgamma} in effective Theory framework  \label{sec:SM}}

The effective weak Lagrangian and SM Wilson coefficients are discussed in section \ref{sec:gen}.
We work out and provide a detailed breakdown of the individual contributions to $D\to V\gamma$ amplitudes in  the heavy-quark approach.
We work out weak annihilation and hard gluon exchange corrections in section \ref{sec:corr}, with
contributions from the gluon dipole operator given in section \ref{sec:Q8}. In section \ref{sec:WA}
we  consider weak annihilation induced modes.

\subsection{Generalities \label{sec:gen}}

The effective $c\to u\gamma$ weak Lagrangian can be written as \cite{deBoer:2015boa}
\begin{align}
 \mathcal L_\text{eff}^\text{weak}=\frac{4G_F}{\sqrt 2}\left(\sum_{q\in\{d,s\}}V_{cq}^*V_{uq}\sum_{i=1}^2C_iQ_i^{(q)}+\sum_{i=3}^6C_iQ_i+\sum_{i=7}^8\left(C_iQ_i+C_i'Q_i'\right)\right)\,,
\end{align}
where $G_F$ is the Fermi constant, $V_{ij}$  are CKM matrix elements and the operators read
\begin{align}
 &Q_1^{(q)}=(\bar u_L\gamma_{\mu_1}T^aq_L)(\overline q_L\gamma^{\mu_1}T^ac_L)\,,&&Q_2^{(q)}=(\bar u_L\gamma_{\mu_1}q_L)(\overline q_L\gamma^{\mu_1}c_L)\,,\nonumber\\
 &Q_3=(\bar u_L\gamma_{\mu_1}c_L)\sum_{\{q:m_q<\mu_c\}}(\overline q\gamma^{\mu_1}q)\,,&&Q_4=(\bar u_L\gamma_{\mu_1}T^ac_L)\sum_{\{q:m_q<\mu_c\}}(\overline q\gamma^{\mu_1}T^aq)\,,\nonumber\\
 &Q_5=(\bar u_L\gamma_{\mu_1}\gamma_{\mu_2}\gamma_{\mu_3}c_L)\sum_{\{q:m_q<\mu_c\}}(\overline q\gamma^{\mu_1}\gamma^{\mu_2}\gamma^{\mu_3}q)\,,&&Q_6=(\bar u_L\gamma_{\mu_1}\gamma_{\mu_2}\gamma_{\mu_3}T^ac_L)\sum_{\{q:m_q<\mu_c\}}(\overline q\gamma^{\mu_1}\gamma^{\mu_2}\gamma^{\mu_3}T^aq)\,,\nonumber\\
 &Q_7=\frac{e\,m_c}{16\pi^2}(\bar u_L\sigma^{\mu_1\mu_2}c_R)F_{\mu_1\mu_2}\,,&&Q_7'=\frac{e\,m_c}{16\pi^2}(\bar u_R\sigma^{\mu_1\mu_2}c_L)F_{\mu_1\mu_2}\,,\nonumber\\
 &Q_8=\frac{g_s\,m_c}{16\pi^2}(\bar u_L\sigma^{\mu_1\mu_2}T^ac_R)G^a_{\mu_1\mu_2}\,,&&Q_8'=\frac{g_s\,m_c}{16\pi^2}(\bar u_R\sigma^{\mu_1\mu_2}T^ac_L)G^a_{\mu_1\mu_2}\,,
\end{align}
where $F_{\mu \nu}, G^a_{\mu \nu}, a=1,..,8$ denote the electromagnetic, gluonic field strength tensor, respectively, and $T^a$  are the generators of QCD.
In the following all Wilson coefficients are understood as evaluated at the charm scale $\mu_c $  of the order of the charm mass $m_c$, and $\mu_c=1.27\,\text{GeV}$ unless 
otherwise explicitly stated.

For the SM Wilson coefficients of $Q_{1,2}$ and the effective coefficient of the chromomagnetic dipole operator
at leading order in $\alpha_s$ one obtains \cite{deBoer:2015boa,deBoer:2016dcg}, respectively, 
\begin{align}
 &C_1^{(0)}\in[-1.28,-0.83]\,,\quad C_2^{(0)}\in[1.14,1.06]\,,\nonumber\\
 &C_8^{(0)\text{eff}}\in[0.47\cdot10^{-5}-1.33\cdot10^{-5}i,0.21\cdot10^{-5}-0.61\cdot10^{-5}]\,,
\end{align}
where $\mu_c$ is varied within $[m_c/\sqrt2,\sqrt2m_c]$.
$C_8^{(0)\text{eff}}$  is strongly GIM suppressed in the SM and negligible therein. $C_1^{(0)}$ and the color suppressed coefficient of  the weak annihilation contribution introduced in section \ref{sec:corr} ,
\begin{align}
 \frac49C_1^{(0)}+\frac13C_2^{(0)}\in[-0.189,-0.018]\,, 
\end{align}
are subject to a large scale-uncertainty. Note, at next-to leading order,  $4/9\,C_1+1/3\,C_2 \in [-0.042,0.092] $. 
In this work first (second) entries in  intervals correspond to the lower (upper) value of $\mu_c$ within $[m_c/\sqrt2,\sqrt2m_c]$.

The effective  coefficient of $Q_7$ including  the matrix elements of $Q_{1-6}$ at two-loop QCD, see  \cite{Greub:1996wn,deBoer:2015boa,deBoer:2016dcg}  and
appendix \ref{app:C7eff} for details, is in the range
\begin{align}  \label{eq:2loop}
 C_7^\text{eff}\in[-0.00151-(0.00556i)_\text{s}+(0.00005i)_\text{CKM},-0.00088-(0.00327i)_\text{s}+(0.00002i)_\text{CKM}]
\end{align}
and $C_7^{\prime\,\text{eff}}\sim m_u/m_c\simeq0$.
Here, we give contributions to the imaginary parts separately: The ones with subscript  "$s$" correspond to strong phases, whereas the ones with label "CKM"
stem from  the weak phases in the CKM matrix.
As a new ingredient  we provide  in this work  the 2-loop QCD matrix element of $Q_{3-6}$, see appendix \ref{app:C7eff} for details.
Numerically,  $C_7^{\text{eff}}\big|_{\langle Q_{3-6}\rangle}\lesssim10^{-6}$, that is,   negligible  due to small SM Wilson coefficients $C_{3-6}$ and the GIM suppression.

The $D\to V\gamma$ decay rate  can be written as \cite{Isidori:2012yx}
\begin{align}\label{eq:Gamma_DtoVgamma}
 \Gamma=\frac{m_D^3}{32\pi}\left(1-\frac{m_V^2}{m_D^2}\right)^3\left(|A_\text{PC}|^2+|A_\text{PV}|^2\right)\,,
\end{align}
where the parity conserving (PC) and parity violating (PV) amplitudes read
\begin{align}
 A_\text{PC/PV}=\frac{\sqrt{\alpha_e4\pi}G_Fm_c}{2\sqrt2\pi^2}(A_7\pm A_7')T
\end{align}
times $1/\sqrt2$ for $V^0\in\{\rho^0,\omega\}$. Here, $m_D$ and $m_V$ denote the mass of the $D$ and the vector meson, respectively, and   $T=T_1(0)=T_2(0)$ is a 
$D \to V$ tensor form factor, see appendix \ref{app:DV_form_factors} for details.
We stress  that the dominant SM contribution  to $D_{(s)} \to V \gamma$ branching ratios is independent of $T$.
Furthermore, 
\begin{align} A_7^{(\prime)}=C_7^{(\prime)\text{eff}}+... \, ,
\end{align}
 where the ellipses indicate additional contributions from  within and outside the SM.
Corrections  from within the SM are obtained in  sections \ref{sec:corr}, \ref{sec:Q8} and \ref{sec:WA}.
BSM coefficients  and amplitudes are denoted by $\delta C$ and $\delta A_7^{(\prime)}$, respectively.
Note, $A_7$ and $A_7^\prime$ do not mix in eq.~(\ref{eq:Gamma_DtoVgamma}).

\subsection{Corrections \label{sec:corr}}

\begin{figure}[!htb]
 \centering
 \includegraphics[trim=2.5cm 22.5cm 2.5cm 3.5cm,clip,width=0.8\textwidth]{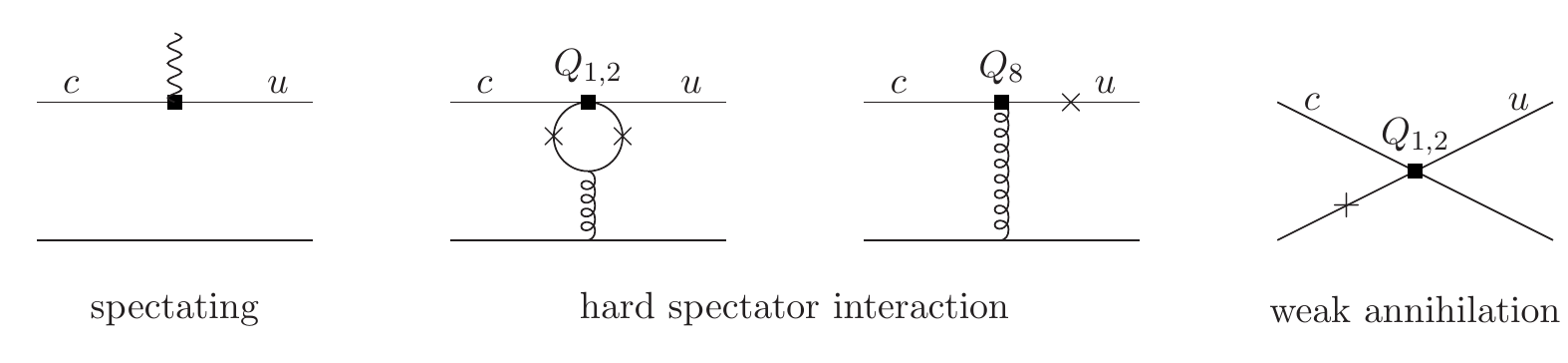}
 \caption{Diagrams driven by  $C_7^{\rm eff}$, weak annihilation and hard spectator interaction.
 The crosses indicate photon emission.
 Diagrams not shown are additionally power suppressed.}
 \label{fig:diagrams}
\end{figure}

In this section we calculate the hard spectator interaction (HSI) and weak annihilation (WA) contributions shown in figure~\ref{fig:diagrams} as corrections to the leftmost  diagram in
the figure.

The leading ($\sim\alpha_s^1\,(\Lambda_\text{QCD}/m_c)^0$) hard spectator interaction within  QCD factorization adopted from $b$-physics \cite{Bosch:2001gv} (also \cite{Ali:2001ez,Beneke:2001at,Beneke:2004dp}) can be written as
\begin{align}\label{eq:C7_HSI}
 C_7^{\text{HSI},V}=\frac{\alpha_s(\mu_h)}{4\pi}\left(\sum_{q\in\{d,s\}}V_{cq}^*V_{uq}\left(-\frac16C_1^{(0)}(\mu_h)+C_2^{(0)}(\mu_h)\right)H_1^{(q)}+C_8^{(0)\text{eff}}(\mu_h)H_8\right)\,,
\end{align}
where we consistently use $C_{1,2,8}^{(0)}$ at leading order in $\alpha_s$ due to additional non-factorizable diagrams at higher order and $\mu_h\sim\sqrt{\Lambda_\text{QCD}m_c}$.
Furthermore,
\begin{align}
 &H_1^{(q)}=\frac{4\pi^2f_Df_V^\perp}{27Tm_D\lambda_D}\int_0^1\mathrm dv\,h_V^{(q)}(\bar v)\,\Phi_{V\perp}(v)\,,\nonumber\\
 &h_V^{(q)}=\frac{4m_q^2}{m_c^2{\bar v}^2}\left(\mathrm{Li_2}\left[\frac2{1-\sqrt{(\bar v-4m_q^2/m_c^2+i\epsilon])/\bar v}}\right]+\mathrm{Li_2}\left[\frac2{1+\sqrt{(\bar v-4m_q^2/m_c^2+i\epsilon])/\bar v}}\right]\right)-\frac2{\bar v}\,,\nonumber\\
 &H_8=-\frac{32\pi^2f_Df_V^\perp}{27Tm_D\lambda_D}\int_0^1\mathrm dv\frac{\Phi_{V\perp}(v)}v\,,
\end{align}
$\bar v=1-v$ and $\mathrm{Li_2}[x]=-\int_0^x\mathrm dt\,\mathrm{ln}[1-t]/t$ and the decay constants $f_D,f_V^\perp$ are given in appendix \ref{app:parameters}.
We use $m_d=0$. As $Q_8^{(\prime)}$-induced HSI contributions are negligible in the SM it follows that $C_7^{\rm HSI}$ is driven by $V_{cs}^* V_{us}$.
The transverse distribution at leading twist is to first order in Gegenbauer polynomials
\begin{align}
 \Phi_{V\perp}=6v\bar v\left(1+a_1^{V\perp}\,3(v-\bar v)+a_2^{V\perp}\,\frac32\left(5(v-\bar v)^2-1\right)\right)\,.
\end{align}
Numerical input on the Gegenbauer moments $a_{1,2}^{V\perp}$ is given  in appendix \ref{app:parameters}.

The parameter $\lambda_D$ is defined as
\begin{align}
 \frac{m_D}{\lambda_D}=\int_0^1\mathrm d\xi\frac{\Phi_D(\xi)}{\xi}\,,
\end{align}
that is the first negative moment of the leading twist distribution amplitude $\Phi_D$ of the light-cone momentum fraction $\xi$ of the spectator quark within the $D$-meson.
In $b$-physics, the first negative moment of the $B$-meson light-cone distribution amplitude, $\lambda_B^\text{HQET}>0.172\,\text{GeV}$ at 90\% C.L. \cite{Heller:2015vvm}, a positive light-cone wave function yields $\lambda_B^\text{HQET}\leq 4/3\,\bar\Lambda$ \cite{Korchemsky:1999qb} and by means of light-cone sum rules (LCSR) $\lambda_B^\text{QCD}\lesssim\bar\Lambda$ \cite{Ball:2003fq,Khodjamirian:2005ea}, where $\bar\Lambda=(m_B-m_b)+\mathcal O(\Lambda_\text{QCD}^2/m_b)$ and $\lambda_B^\text{HQET}>\lambda_B^\text{QCD}$ at one-loop QCD \cite{Pilipp:2007sb}.
We use $\lambda_D\sim\Lambda_\text{QCD}\sim\mathcal O(0.1\,\text{GeV})$.

Taking $\mu_h=1\,\text{GeV}$, varying the Gegenbauer moments and decay constants (but not the form factor $T$ as it cancels in the amplitude) we find
\begin{align} \label{eq:HSI}
 &C_7^{\text{HSI},\rho}\in[0.00051+0.0014i,0.00091+0.0020i]\cdot\frac{\text{GeV}}{\lambda_D}\,,\nonumber\\
 &C_7^{\text{HSI},\omega}\in[0.00030+0.0010i,0.00098+0.0020i]\cdot\frac{\text{GeV}}{\lambda_D}\,,\nonumber\\
 &C_7^{\text{HSI},{K^*}^+}\in[0.00032+0.0013i,0.00096+0.0022i]\cdot\frac{\text{GeV}}{\lambda_D}\,.
\end{align}
We neglect isospin breaking in the Gegenbauer moments of the $\rho$. Contributions induced by $Q_8^{(\prime)}$ are discussed in section \ref{sec:Q8}.

The leading ($\sim\alpha_s^0\,(\Lambda_\text{QCD}/m_c)^1$) weak annihilation contribution to  $D^0\to(\rho^0,\omega)\gamma$, $D^+\to\rho^+\gamma$ and $D_s\to{K^*}^+\gamma$ can be inferred from $b$-physics \cite{Bosch:2001gv,Bosch:2004nd}. We obtain
\begin{align}
 &C_7^{\text{WA},\rho^0}=-\frac{2\pi^2Q_uf_Df_{\rho^0}^{(d)}m_\rho}{Tm_{D^0}m_c\lambda_D}V_{cd}^*V_{ud}\left(\frac49C_1^{(0)}+\frac13C_2^{(0)}\right)\,,\nonumber\\
 &C_7^{\text{WA},\omega}=\frac{2\pi^2Q_uf_Df_\omega^{(d)}m_\omega}{Tm_{D^0}m_c\lambda_D}V_{cd}^*V_{ud}\left(\frac49C_1^{(0)}+\frac13C_2^{(0)}\right)\,,\nonumber\\
 &C_7^{\text{WA},\rho^+}=\frac{2\pi^2Q_df_Df_\rho m_\rho}{Tm_{D^+}m_c\lambda_D}V_{cd}^*V_{ud}\,C_2^{(0)}\,,\nonumber\\
 &C_7^{\text{WA},{K^*}^+}=\frac{2\pi^2Q_df_{D_s}f_{K^*}m_{K^*}}{Tm_{D_s}m_c\lambda_D}V_{cs}^*V_{us}\,C_2^{(0)}\,,
\label {eq:C7_WAanal}
\end{align}
where $Q_{u}=2/3$, $Q_d=-1/3$ and  we consistently  use $C_{1,2}^{(0)}$ at leading order in $\alpha_s$.
We neglect weak annihilation contributions from  $Q_{3-6}$ as the corresponding Wilson coefficients in the SM are strongly GIM suppressed.
The minus sign for $\rho^0$ is due to isospin.

Varying the decay constants and $\mu_c$ within $[m_c/\sqrt2,\sqrt2m_c]$ we find
\begin{align}\label{eq:C7_WA}
 &C_7^{\text{WA},\rho^0}\in[-0.010,-0.0011]\cdot\frac{\text{GeV}}{\lambda_D}\,,\nonumber\\
 &C_7^{\text{WA},\omega}\in[0.0097,0.0011]\cdot\frac{\text{GeV}}{\lambda_D}\,,\nonumber\\
 &C_7^{\text{WA},\rho^+}\in[0.029,0.038]\cdot\frac{\text{GeV}}{\lambda_D}\,,\nonumber\\
 &C_7^{\text{WA},{K^*}^+}\in[-0.034,-0.047]\cdot\frac{\text{GeV}}{\lambda_D}\,.
\end{align}
Note that non-factorizable power corrections (inducing $A_7'$) could in principle be calculated with LCSR, see, {\it  e.g.,} \cite{Ball:2006eu} and that non-local corrections to weak annihilation by means of QCD sum rules are additionally power suppressed \cite{Kagan:2001zk}.

To summarize,  we observe the following hierarchies among the SM contributions to $A_7$
\begin{align} \label{eq:hi}
|C_7^{\text{WA},V^+}|>|C_7^{\text{WA},V^0}|  \gtrsim  |C_7^{\text{HSI}}|>|C_7^{\text{eff}}|  \, . 
\end{align}
 The leading SM uncertainties are therefore those stemming from the WA-amplitudes, that is, the  $\mu_c$-scale and $\lambda_D$ uncertainties, followed by the parameters entering HSI-amplitudes, {\it i.e.}, Gegenbauer moments, decay constants and the $\mu_h$-scale. The latter we fixed for simplicity.

 Contributions to  $A_7'$ arise in the SM from  a $Q_2$-induced quark loop with a soft gluon  as a power correction \cite{Grinstein:2004uu}
 \begin{align} \label{eq:soft}
 C_7^{\prime(c\to u\gamma g)}\sim\frac13C_2^{(0)}\left(V_{cd}^*V_{ud}\,f^{(c\to u\gamma g)}(m_d^2/m_c^2)+V_{cs}^*V_{us}\,f^{(c\to u\gamma g)}(m_s^2/m_c^2)\right)\frac{\Lambda_\text{QCD}}{m_c} \, , 
\end{align}
which is $\mathcal O(10^{-4})$
if the expansion coefficients of $f^{(c\to u\gamma g)}$ in $m_q^2/m_c^2$ are order one. 
Note that the $c\to u\gamma g$ process induces as well a contribution to $A_7$, and that the $Q_1$-induced quark loop is additionally color suppressed.
Note also that $f^{(c\to u\gamma g)}$ could in principle be calculated with LCSR, see {\it e.g.,} \cite{Ball:2006eu}, yet $\alpha_s$-corrections vanish at leading twist in the limit of massless quarks in the $V$-meson \cite{Grinstein:2000pc}.
 To be specific, and in absence of further calculations, we limit  the size of the chirality-flipped SM amplitudes  in our numerical analysis as
 \begin{align} \label{eq:A7prime}
| A_{7,\rm SM}'/A_{7,\rm SM}|\lesssim0.2 \, ,
 \end{align}
 and take the structure of the weak phases as  in (\ref{eq:soft}) into account.

\subsection{Contributions from $Q_8^{(\prime)}$ \label{sec:Q8}}

We detail here the  contributions from $Q_8^{(\prime)}$ to $c \to u \gamma$ modes. While in the 
 SM  they  are negligibly small they can be relevant in BSM scenarios.
 
Numerically, we find for  the $\langle Q_8^{(\prime)}\rangle$-induced  hard spectator interaction of eq.~(\ref{eq:C7_HSI})
\begin{align}
 &C_7^{(\prime)\text{HSI},\rho}\big|_{\langle Q_8^{(\prime)}\rangle}\in-\frac{\text{GeV}}{\lambda_D}\cdot[0.031,0.042]\cdot C_8^{(\prime)}\,,\nonumber\\
 &C_7^{(\prime)\text{HSI},\omega}\big|_{\langle Q_8^{(\prime)}\rangle}\in-\frac{\text{GeV}}{\lambda_D}\cdot[0.024,0.040]\cdot C_8^{(\prime)}\,,\nonumber\\
 &C_7^{(\prime)\text{HSI},{K^*}^+}\big|_{\langle Q_8^{(\prime)}\rangle}\in-\frac{\text{GeV}}{\lambda_D}\cdot[0.031,0.039]\cdot C_8^{(\prime)} \, . \label{eq:C7_Q8_HSI}
\end{align}
 Note that QCD factorization breaks down at subleading power for $\langle Q_8\rangle$ hard spectator interaction due to a logarithmic singularity for a soft spectator quark \cite{Kagan:2001zk}.  

Alternatively, LCSR yield the $\langle Q_8^{(\prime)}\rangle$ gluon spectator interaction (GSI) \cite{Dimou:2012un}
\begin{align}
 &C_7^{(\prime)\text{GSI},\rho^0}\in-[0.068+0.048i,0.14+0.10i]\cdot C_8^{(\prime)}\,,\nonumber\\
 &C_7^{(\prime)\text{GSI},\omega}\in-[0.018-0.024i,0.036-0.048i]\cdot C_8^{(\prime)}\,,\nonumber\\
 &C_7^{(\prime)\text{GSI},\rho^+}\in-[0.057+0.040i,0.12+0.083i]\cdot C_8^{(\prime)}\,,\nonumber\\
 &C_7^{(\prime)\text{GSI},{K^*}^+}\in-[0.017-0.020i,0.034-0.040i]\cdot C_8^{(\prime)} \, . \label{eq:C7_Q8_GSI}
\end{align}
The contributions in eqs.~(\ref{eq:C7_Q8_GSI}) and (\ref{eq:C7_Q8_HSI}) are similar in size for $\lambda_D\sim\mathcal O(0.1\,\text{GeV})$.
One may  compare these to  the $\langle Q_8^{(\prime)}\rangle$  induced  contribution to eq.~(\ref{eq:2loop})
\begin{align}\label{eq:C7_Q8_eff}
 C_7^{(\prime)\text{eff}}\big|_{\langle Q_8^{(\prime)}\rangle}\simeq(-0.12-0.17i)C_8^{(\prime)}\,.
\end{align}
BSM values  $\delta C_8^{(\prime)} \lesssim  0.1$ 
can therefore  lift the HSI/GSI contributions and the one to $C_7^{(\prime)\text{eff}}$ such that
\begin{align} \label{eq:hiBSM}
|C_7^{\text{WA},V^+}|>|C_7^{\text{WA},V^0}|  \gtrsim  |C_7^{\text{HSI}}|, |C_7^{\text{eff}}|  \,.
\end{align}

\subsection{Weak annihilation induced modes \label{sec:WA}}

The contributions to $A_7$ of the  weak annihilation induced decays $D^0\to(\phi,\bar K^{*0},{K^*}^0)\gamma$, $D^+\to{K^*}^+\gamma$ and $D_s\to\rho^+\gamma$ are obtained as
follows
\begin{align}
 &C_7^{D^0\to\phi\gamma}=\frac{2\pi^2Q_uf_Df_\phi m_\phi}{Tm_{D^0}m_c\lambda_D}V_{cs}^*V_{us}\left(\frac49C_1^{(\bar us)(\bar sc)}+\frac13C_2^{(\bar us)(\bar sc)}\right)\,,\nonumber\\
 &C_7^{D^0\to\bar K^{*0}\gamma}=\frac{2\pi^2Q_uf_Df_{K^*}m_{{K^*}^0}}{Tm_{D^0}m_c\lambda_D}V_{cs}^*V_{ud}\left(\frac49C_1^{(\bar ud)(\bar sc)}+\frac13C_2^{(\bar ud)(\bar sc)}\right)\,,\nonumber\\
 &C_7^{D^0\to{K^*}^0\gamma}=\frac{V_{cd}^*V_{us}}{V_{cs}^*V_{ud}}C_7^{D^0\to\bar K^{*0}\gamma}\,,\nonumber\\
 &C_7^{D^+\to{K^*}^+\gamma}=\frac{2\pi^2Q_df_Df_{K^*}m_{{K^*}^+}}{Tm_{D^+}m_c\lambda_D}V_{cd}^*V_{us}C_2^{(\bar us)(\bar dc)}\,,\nonumber\\
 &C_7^{D_s\to\rho^+\gamma}=\frac{2\pi^2Q_df_{D_s}f_\rho m_\rho}{Tm_{D_s}m_c\lambda_D}V_{cs}^*V_{ud}C_2^{(\bar ud)(\bar sc)}\,.
 \label{eq:pureWA}
\end{align}
Here we made the flavor structure of the Wilson coefficients explicit,  however,  since QCD is flavor symmetric, use
$C_{1/2}^{(\bar q_1q_2)(\bar q_3c)}=C_{1/2}^{(0)}$.
While the form factor $T$ is process-dependent, it cancels together with $m_c$ in the decay amplitude.
Numerically, $\text{GeV}/(m_cT)\sim1$ .
Varying the decay constants and $\mu_c\in[m_c/\sqrt2,\sqrt2m_c]$ we find
\begin{align}
 &C_7^{D^0\to\phi\gamma}\in[-0.016,-0.0013]\cdot\frac{\text{GeV}}{\lambda_D}\frac{\text{GeV}}{m_cT}\,,\nonumber\\
 &C_7^{D^0\to\bar K^{*0}\gamma}\in[-0.051,-0.0044]\cdot\frac{\text{GeV}}{\lambda_D}\frac{\text{GeV}}{m_cT}\,,\nonumber\\
 &C_7^{D^0\to{K^*}^0\gamma}\in[0.0028,0.00023]\cdot\frac{\text{GeV}}{\lambda_D}\frac{\text{GeV}}{m_cT}\,,\nonumber\\
 &C_7^{D^+\to{K^*}^+\gamma}\in[0.0082,0.0070]\cdot\frac{\text{GeV}}{\lambda_D}\frac{\text{GeV}}{m_cT}\,,\nonumber\\
 &C_7^{D_s\to\rho^+\gamma}\in[-0.16,-0.13]\cdot\frac{\text{GeV}}{\lambda_D}\frac{\text{GeV}}{m_cT}\,.
\end{align}

For $D^0 \to \phi \gamma$ additional contributions to the decay amplitude can arise, induced by 
 $d \bar d + u \bar u$-admixture in the $\phi$ or rescattering \cite{Isidori:2012yx}. Such effects can be  parametrized by  $y$ as follows
\begin{align}
{\cal{A}}(D^0 \to \phi \gamma)\simeq V_{cs}^* V_{us} a^{\rm WA}_\phi + y  \left[   V_{cd}^* V_{ud} \,  a^{\rm WA}_{\rho^0} -V_{cs}^* V_{us} \,  a^{\rm HSI}_{\rho^0} -a^{C_7^{\rm eff}}  \right] \, .
\end{align}
To estimate $A_{CP}$ we made CKM factors explicit except for the $C_7^{\rm eff}$-induced term which can receive large BSM CP violating phases.
The amplitudes correspond, in order of appearance, to 
$C_7^{D^0\to\phi\gamma}$,  and the three contributions eqs.~(\ref{eq:C7_WAanal}), (\ref{eq:HSI}) and (\ref{eq:2loop}) to $D^0 \to \rho^0 \gamma$. Note the minus signs due do the $SU(3)$-composition.
One obtains, model-independently,
\begin{align} \label{eq:ACPphi}
A_{CP} (D^0 \to \phi \gamma) \simeq |y| A_{CP} (D^0 \to \rho^0\gamma)  + {\cal{O}}(y^2) \, . 
\end{align}

\section{SM Phenomenology  \label{sec:pheno}}

We provide SM predictions for various $D_{(s)} \to V \gamma$ modes and compare to existing data.
In addition to the QCD-based approach of the previous  section we present branching ratios  in the phenomenological  approach of  \cite{Fajfer:1997bh,Fajfer:1998dv}.
This model is a hybrid of factorization, heavy quark effective  and chiral  perturbation theory, where the $SU(3)$ flavor symmetry is broken by  measured parameters.
Compared to \cite{Fajfer:1997bh,Fajfer:1998dv} we rewrite the amplitudes in terms of newly measured parameters and vary (updated) parameters within uncertainties.
Analytical expressions for the   $D\to V\gamma$ amplitudes are provided  in appendix \ref{app:resonant_DVgamma_amplitudes}.
The hierarchies of the various $D\to V\gamma$ amplitudes are predominantly set by  CKM factors and large-$N_C$ counting,
taken care of  in both the heavy quark and the hybrid frameworks.

The SM branching ratios  and presently available data are given in table~\ref{tab:DVgamma_branching_ratios}.
\begin{table}[!htb]
 \centering
 \begin{tabular}{c|c|c|c|c}
  branching ratio                                                                              &  $D^0\to\rho^0\gamma$         &  $D^0\to\omega\gamma$        &  $D^+\to\rho^+\gamma$        &  $D_s\to{K^*}^+\gamma$  \\
  \noalign{\hrule height 1pt}
  two-loop QCD                                                                                 &  $(0.14-2.0)\cdot10^{-8}$     &  $(0.14-2.0)\cdot10^{-8}$    &  $(0.75-1.0)\cdot10^{-8}$    &  $(0.32-5.5)\cdot10^{-8}$   \\
  HSI+WA                                                                                       &  $(0.11-3.8)\cdot10^{-6}$     &  $(0.078-5.2)\cdot10^{-6}$   &  $(1.6-1.9)\cdot10^{-4}$     &  $(1.0-1.4)\cdot10^{-4}$  \\
  \hline
  hybrid                                                                                       &  $(0.041-1.17)\cdot10^{-5}$   &  $(0.042-1.12)\cdot10^{-5}$  &  $(0.017-2.33)\cdot10^{-4}$  &  $(0.053-1.54)\cdot10^{-4}$  \\
  \hline
  \cite{Fajfer:1997bh,Fajfer:1998dv}                                                           &  $(0.1-1)\cdot10^{-5}$        &  $(0.1-0.9)\cdot10^{-5}$     &  $(0.4-6.3)\cdot10^{-5}$     &  $(1.2-5.1)\cdot10^{-5}$  \\
  \cite{Burdman:1995te}                                                                        &  $(0.1-0.5)\cdot10^{-5}$      &  $0.2\cdot10^{-5}$           &  $(2-6)\cdot10^{-5}$         &  $(0.8-3)\cdot10^{-5}$  \\
  \cite{Khodjamirian:1995uc}\footnote{Uncertainties not available. We take $a_1=1.3$ and $a_2=-0.55$ \cite{Bauer:1986bm}.}  &  $3.8\cdot10^{-6}$            &  --                          &  $4.6\cdot10^{-6}$           &  --  \\
  \hline
  data$^\dagger$                                                                  &  $(1.77\pm0.31)\cdot10^{-5}$  &   $<2.4\cdot10^{-4}$         &  --                          &  --
 \end{tabular}
 \caption{Branching ratios of $D\to V\gamma$ within the SM at two-loop QCD, from the  hard spectator interaction plus weak annihilation and the hybrid approach.
 We vary the form factors, decay constants, lifetimes, Gegenbauer moments, relative strong phases and $\mu_c\in[m_c/\sqrt2,\sqrt2m_c]$.
 The branching ratios from the  hard spectator interaction plus weak annihilation scale as $(0.1\,\text{GeV})/\lambda_D)^2$.
 Also given  are data by  the Belle  \cite{Abdesselam:2016yvr} and the CLEO (at 90\% CL) \cite{Asner:1998mv} collaborations as well as  SM predictions  from \cite{Fajfer:1997bh,Fajfer:1998dv}, via pole diagrams and  VMD  \cite{Burdman:1995te} and QCD sum rules \cite{Khodjamirian:1995uc}. $^\dagger$Statistical and systematic uncertainties are added in quadrature.}
 \label{tab:DVgamma_branching_ratios}
\end{table}
We learn the following: The branching ratios induced by  hard spectator interaction plus weak annihilation are typically smaller than (similar to) the ones obtained in the hybrid approach for neutral (charged) $c\to u\gamma$ modes. The branching ratio from two-loop QCD eq.~(\ref{eq:2loop}) is subleading in each case.
The  branching ratios in the hybrid approach cover the ranges  previously obtained  in \cite{Burdman:1995te,Khodjamirian:1995uc,Fajfer:1997bh,Fajfer:1998dv}.
The measured $D^0\to\rho^0\gamma$ branching ratio is somewhat above the SM prediction in the hybrid model.

\begin{figure}[!htb]
 \centering
 \includegraphics[width=0.8\textwidth]{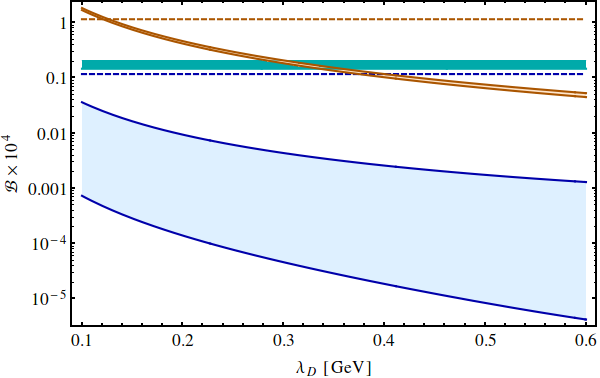}
 \caption{Branching ratios of $D\to\rho\gamma$ as a function of $\lambda_D$.
 The upper orange curves are for $D^+\to\rho^+\gamma$ and the lower blue curves are for $D^0\to\rho^0\gamma$.
 The solid curves represent the bands of two-loop QCD and hard spectator interaction plus weak annihilation, the dashed lines are the maximal predictions in the hybrid approach and the cyan band depicts the measured branching ratio \cite{Abdesselam:2016yvr}.
 We vary the form factors, decay constants, lifetimes, Gegenbauer moments, relative strong phases and $\mu_c\in[m_c/\sqrt2,\sqrt2m_c]$.}
 \label{fig:BlambdaDDrhogamma}
\end{figure}

The branching ratios of $D\to\rho\gamma$ as a function of $\lambda_D$ are shown in figure~\ref{fig:BlambdaDDrhogamma}.
The $D^+\to\rho^+\gamma$ SM branching ratio is  $\lesssim2\cdot10^{-4}$, a measurement  would constrain $\lambda_D$ efficiently.
 Specifically,  we find  $\mathcal B(D^+\to\rho^+\gamma)\simeq[44,2900]\cdot\mathcal B(D^0\to\rho^0\gamma)$ by means of hard spectator interaction plus weak annihilation and in the hybrid  model $\mathcal B (D^+\to\rho^+\gamma)\simeq[0.3,280]\cdot\mathcal B(D^0\to\rho^0\gamma)$. The $D^0\to\rho^0\gamma$ branching ratio can be subject to stronger cancellations between the contributions in eq.~(\ref{eq:hi}) than in the hybrid model.
Assuming that the phase of each amplitude $A_\text{PV/PC}^\text{I/II/III}$ is equal for $D^+\to\rho^+\gamma$ and $D^0\to\rho^0\gamma$ reduces the possible isospin breaking to  $\mathcal B (D^+\to\rho^+\gamma)\simeq[0.6,140]\cdot\mathcal B (D^0\to\rho^0\gamma)$.
Note, isospin is already significantly broken by the  lifetimes $\tau(D^0)/\tau(D^+) \simeq0.4$  \cite{Olive:2016xmw}.

The uncertainties in the hybrid model are dominated by the relative strong phases, followed by the phenomenological fit  coefficients $a_1=1.3\pm0.1$, $a_2=-0.55\pm0.1$ \cite{Bauer:1986bm} (also \cite{Buras:1994ij,Cheng:2002wu}).

The branching ratios of $D^0\to(\phi,\bar K^{*0},{K^*}^0)\gamma$, $D^+\to{K^*}^+\gamma$ and $D_s\to\rho^+\gamma$ are given in 
table~\ref{tab:Dphigamma_DKstar0gamma_branching_ratios}.
\begin{table}[!htb]
 \centering
 \begin{tabular}{c|c|c|c|c|c}
  branching ratio                                                                              &  $D^0\to\phi\gamma$           &  $D^0\to\bar K^{*0}\gamma$    &  $D^0\to{K^*}^0\gamma$      &  $D^+\to{K^*}^+\gamma$      &  $D_s\to\rho^+\gamma$  \\
  \noalign{\hrule height 1pt}
  WA                                                                                           &  $(0.0074-1.2)\cdot10^{-5}$   &  $(0.011-1.6)\cdot10^{-4}$    &  $(0.032-4.4)\cdot10^{-7}$  &  $(0.73-1.1)\cdot10^{-5}$   &  $(1.8-2.9)\cdot10^{-3}$  \\
  \hline
  hybrid                                                                                       &  $(0.24-2.8)\cdot10^{-5}$     &  $(0.26-4.6)\cdot10^{-4}$     &  $(0.076-1.3)\cdot10^{-6}$  &  $(0.48-7.6)\cdot10^{-6}$   &  $(0.11-1.3)\cdot10^{-3}$  \\
  \hline
  \cite{Fajfer:1997bh,Fajfer:1998dv}                                                           &  $(0.4-1.9)\cdot10^{-5}$      &  $(6-36)\cdot10^{-5}$         &  $(0.03-0.2)\cdot10^{-5}$   &  $(0.03-0.44)\cdot10^{-5}$  &  $(20-80)\cdot10^{-5}$  \\
  \cite{Burdman:1995te}                                                                        &  $(0.1-3.4)\cdot10^{-5}$      &  $(7-12)\cdot10^{-5}$         &  $0.1\cdot10^{-6}$         &  $(0.1-0.3)\cdot10^{-5}$    &  $(6-38)\cdot10^{-5}$  \\
  \cite{Khodjamirian:1995uc}\footnote{Uncertainties not available. We use  $a_1=1.3$ and $a_2=-0.55$ \cite{Bauer:1986bm}. }  &  --                           &  $1.8\cdot10^{-4}$            &  --                         &  --                         &  $4.7\cdot10^{-5}$  \\
  \hline
  Belle \cite{Abdesselam:2016yvr}$^\dagger$                                                             &  $(2.76\pm0.21)\cdot10^{-5}$  &  $(4.66\pm0.30)\cdot10^{-4}$  &  --                         &  --                         &  --  \\
  BaBar \cite{Aubert:2008ai}$^\dagger$\footnote{We update the normalization \cite{Olive:2016xmw}.}       &  $(2.81\pm0.41)\cdot10^{-5}$  &  $(3.31\pm0.34)\cdot10^{-4}$  &  --                         &  --                         &  --
 \end{tabular}
 \caption{Branching ratios of $D^0\to(\phi,\bar K^{*0},{K^*}^0)\gamma$, $D^+\to{K^*}^+\gamma$ and $D_s\to\rho^+\gamma$ within the SM from weak annihilation and within the hybrid   framework \cite{Fajfer:1997bh,Fajfer:1998dv} (appendix \ref{app:resonant_DVgamma_amplitudes}).
 We vary the decay constants, lifetimes and $\mu_c\in[m_c/\sqrt2,\sqrt2m_c]$.
 The branching ratios induced by  weak annihilation scale as $(0.1\,\text{GeV})/\lambda_D)^2$. Also given  are available data by 
 the Belle  \cite{Abdesselam:2016yvr} and BaBar \cite{Aubert:2008ai} collaborations, as well as SM predictions obtained in \cite{Fajfer:1997bh,Fajfer:1998dv}, via pole diagrams and  VMD  \cite{Burdman:1995te} and QCD sum rules \cite{Khodjamirian:1995uc}.
 $^\dagger$Statistical and systematic uncertainties are added in quadrature.}
 \label{tab:Dphigamma_DKstar0gamma_branching_ratios}
\end{table}
The measurements by Belle \cite{Abdesselam:2016yvr} and BaBar   \cite{Aubert:2008ai} of  ${\cal{B}}(D^0\to\bar K^{*0}\gamma)$ differ by $2.2\sigma$, yet both are in the range of the
hybrid  model predictions.
Interpreted in the QCD framework to the order we are working, ${\cal{B}}(D^0\to (\bar K^{*0}, \phi) \gamma)$ data
require  a low value of $\lambda_D$ below 0.1 GeV or a  low charm mass scale $\mu_c \sim m_c/2$,  similar to ${\cal{B}}(D^0 \to \rho^0 \gamma)$ data assuming the SM.
Quite generally the deficiency in explaining the largely SM-dominated branching ratios in table~\ref{tab:Dphigamma_DKstar0gamma_branching_ratios}
suggest slow convergence of the $1/m_D$, $\alpha_s$-expansion. Measurements of branching ratios of  color allowed  $D^+, D_s$ mesons could shed light on this, as here the scale uncertainty is smaller.

\begin{figure}[!htb]
 \centering
 \includegraphics[width=0.8\textwidth]{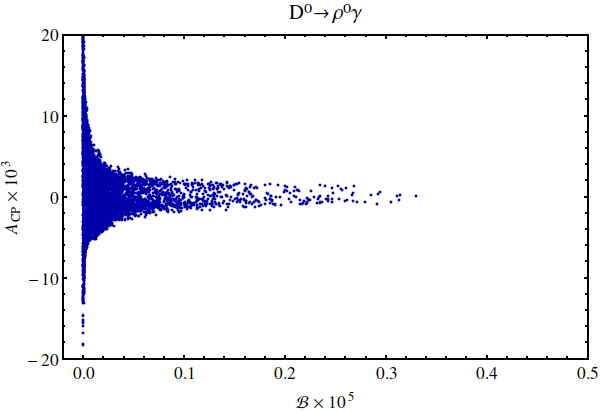}
 \caption{The CP asymmetry versus the branching ratio  for $D^0\to\rho^0\gamma$ decays in the SM.
 We vary the form factor, the two-loop QCD and hard spectator interaction plus weak annihilation within uncertainties, where $\lambda_D\in[0.1,0.6]\,\text{GeV}$, $A_7^\prime$-contributions as in eq.~(\ref{eq:A7prime}) and  relative strong phases.
 The measured $A_{CP}$  eq.~(\ref{eq:ACP}) covers the shown range, whereas the measured branching ratio at one $\sigma$ is above it.}
 \label{fig:ACPB}
\end{figure}

The CP asymmetry for $D^0 \to \rho^0 \gamma$ within the SM in the heavy quark-based approach  is shown in figure~\ref{fig:ACPB} as a function of the SM $D^0 \to \rho^0 \gamma$   branching ratio.
Within the SM $|A_{CP}|\lesssim2\cdot10^{-2}$, if the  branching ratio is $\gtrsim10^{-9}$ and $|A_{CP}|\lesssim2\cdot10^{-3}$, if $\mathcal B (D^0  \to  \rho^0 \gamma) \gtrsim10^{-6}$, {\it e.g.,} for  $\lambda_D\lesssim0.3$ GeV, and as measured (\ref{eq:ACP}) assuming the SM.
$A_{CP}^{\rm SM}$ calculated in the hybrid model is $\lesssim10^{-3}$, and vanishes in the $SU(3)$-limit.
The SM CP asymmetry for $D^0\to\omega\gamma$ is very similar to $A_{CP}(D^0\to\rho^0\gamma)$; the CP asymmetries in the SM for the charged decays  $D^+\to\rho^+\gamma$ and $D_s\to{K^*}^+\gamma$ are $\lesssim2\cdot10^{-3}$ and $\lesssim3\cdot10^{-4}$ in the heavy quark-based approach and the hybrid model, respectively. The enhancement of $A_{CP}$ possible together  with very low values of the branching ratio for $D^0$ decays originates from cancellations 
between different amplitudes.

The SM CP asymmetries for the pure WA modes $D^0\to(\bar K^{*0},{K^*}^0)\gamma$, $D^+\to{K^*}^+\gamma$ and $D_s\to\rho^+\gamma$ vanish  due to a single weak phase {\it c.f.} eq.~(\ref{eq:pureWA}) and appendix \ref{app:resonant_DVgamma_amplitudes}. 
The decay $D^0 \to \phi \gamma$ is special as it can receive contributions at a fraction $y$ similar to  $D^0 \to \rho^0 \gamma$ decays, and has therefore a finite CP asymmetry, estimated in equation (\ref{eq:ACPphi}).
Taking into   account  a percent level $ u \bar u +d \bar d$ content in the $\phi$ \cite{Olive:2016xmw} values 
 of $A_{CP}$ up to 
 $ {\cal{O}}(10^{-4})$ in the SM and  up to $ {\cal{O}}(10^{-3})$ in BSM models can arise in $D^0\to\phi\gamma$ decays.  
 {}Effects from rescattering  at the $\phi$-mass are roughly $y \lesssim 0.1$, hence corresponding CP asymmetries 
 can reach $ {\cal{O}}(10^{-3})$ in the SM and $ {\cal{O}}(10^{-2})$ in BSM scenarios.
The following  asymmetries have been measured   \cite{Abdesselam:2016yvr},
\begin{align} \label{eq:WACP}
A_{CP}(D^0\to\phi\gamma)=-0.094\pm0.066 \pm 0.001 \, , \quad A_{CP}(D^0\to\bar K^{*0}\gamma)=-0.003\pm0.020 \pm 0.000 \, .
\end{align}
$A_{CP}(D^0\to\phi\gamma) $ exhibits presently a mild tension with zero.  

We stress that in our numerical evaluations we vary  all relative strong (unknown) phases, including those   between the WA+HS contributions
and the perturbative ones.  In view of the  appreciable uncertainties we refrain from   putting  an exact upper limit  on the SM-induced CP asymmetries, but  consider, to be specific,  CP asymmetries at percent-level and higher  as an indicator of BSM physics, consistent with \cite{Isidori:2012yx}.
This is supported by the large measured branching fractions, which indicate unsuppressed WA topologies. For the FCNC decays this  suggests no large cancellations between the contributions in eq.~(\ref{eq:hi}), allowing for possible
additional suppressions of CP asymmetries beyond CKM factors.

\section{\texorpdfstring{$D\to V\gamma$}{DtoVgamma} beyond the Standard Model \label{sec:BSM}}

In section \ref{sec:model_independently} we work out model-independent  constraints on  $A_7^{(\prime)}$, $C_7^{(\prime)}$ and $C_8^{(\prime)}$. 
We calculate  BSM Wilson coefficients within leptoquark models in section \ref{sec:leptoquarks} and in SUSY in section \ref{sec:smia}, respectively, and discuss BSM implications.

\subsection{Model-independently}\label{sec:model_independently}

Model-independently,  from  $\mathcal B(D^0\to\rho^0\gamma)$ data, eq.~(\ref{eq:ACP}), we obtain 
\begin{align}  \label{eq:limit}
|A_7^{(\prime)},\delta A_7^{(\prime)}| \lesssim 0.5\, .
\end{align}
Constraints from $\mathcal B(D^+\to\pi^+\mu^+\mu^-)$ data are similar \cite{deBoer:2015boa}.
These constraints prohibit that decays  $D^+\to\rho^+\gamma$ and $D_s\to{K^*}^+\gamma$ are dominated by a BSM dipole contribution. 
Still, a sizable $\delta C_7^{(\prime)}$ can give the leading contribution to the neutral modes, causing their branching ratios to be very close to each other,
  $\mathcal B(D^0\to \omega \gamma) \sim \mathcal B(D^0\to \rho^0 \gamma)$, similar to the  SM,
see table~\ref{tab:DVgamma_branching_ratios}.
Non-observation of this  correlation indicates the presence of intermediate values of $\delta C_7^{(\prime)}$ \cite{Fajfer:2000zx}.

BSM-induced CP asymmetries  can reach $\mathcal O(0.1)$. Hence, current data on $A_{CP}(D^0\to\rho^0\gamma)$  are too uncertain to provide  further constraints. 
There is  essentially no BSM pattern for CP asymmetries  apart from their sign. Turning this   around, it is possible to have a sizable value of $A_{CP}$ in one mode but a small one in another.

\begin{figure}[!htb]
 \centering
 \includegraphics[width=0.48\textwidth]{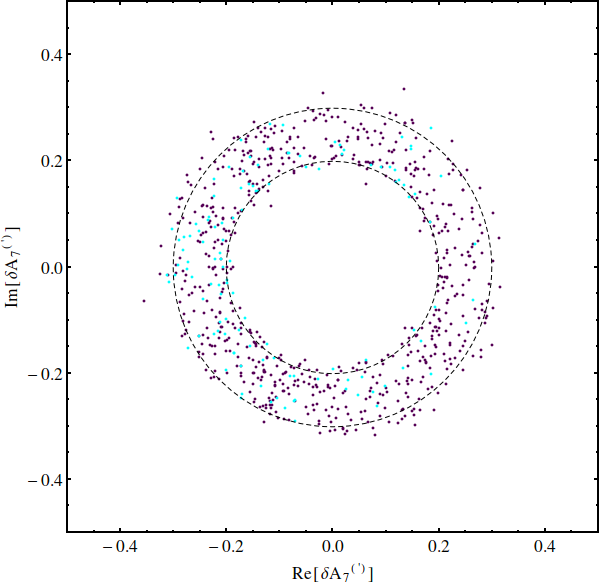}
 \hspace{1em}
 \includegraphics[width=0.48\textwidth]{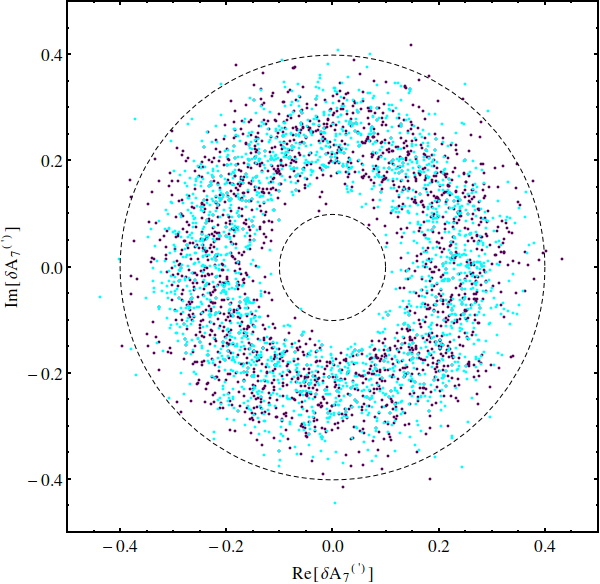}
 \caption{Projected model-independent constraints on $(\mathrm{Re}[\delta A_7],\mathrm{Im}[\delta A_7])$ (purple) and $(\mathrm{Re}[\delta A_7'],\mathrm{Im}[\delta A_7'])$ (cyan), 
 from eq.~(\ref{eq:future})
 for  the heavy quark-based approach (left plot) and the hybrid model  (right plot).
 The dashed circles show $|\delta A_7^{(\prime)}|=0.2,0.3$ (left plot) and $|\delta A_7^{(\prime)}|=0.1,0.4$ (right plot) to guide the eye.
 We vary the form factor, the two-loop QCD, the hard spectator interaction plus weak annihilation coefficients, where $\lambda_D\in[0.1,0.6]\,\text{GeV}$ and  eq.~(\ref{eq:A7prime})  (left plot) and amplitudes in the hybrid model (right plot) within uncertainties and relative phases.}
 \label{fig:C7BSM}
\end{figure}

To illustrate the impact of  improved measurements of the $D^0 \to \rho^0 \gamma$ branching ratio and CP asymmetry, we assume
hypothetical data with a factor four reduced  statistical uncertainty of the current measurements with central values kept \cite{Abdesselam:2016yvr}, that is, 
\begin{align} \label{eq:future}
A_{CP}^{(\rm hypothetical)}(D^0\to\rho^0\gamma)=0.056\pm0.038,\, ~  \mathcal B^{(\rm hypothetical)}(D^0\to\rho^0\gamma)=(1.77\pm0.10)\cdot10^{-5} \, .
\end{align}
In figure~\ref{fig:C7BSM} corresponding 1 sigma-constraints are shown. The ones for $A_7$ (purple) and $A_7^\prime$ (cyan) are similar.
The bounds are stronger for the QCD-based framework,  roughly $0.2\lesssim|\delta A_7^{(\prime)}|\lesssim0.3$, $|\mathrm{Im}[\delta A_7^{(\prime)}]|\gtrsim\mathcal O(0.001)$ (plot to the left) than for  the hybrid one,  $0.1\lesssim|\delta A_7^{(\prime)}|\lesssim0.4$, $|\mathrm{Im}[\delta A_7]|\gtrsim\mathcal O(0.001)$, $|\mathrm{Im}[\delta A_7']|\gtrsim\mathcal O(0.0001)$ (plot to the right), improving on the present situation, eq.~(\ref{eq:limit}).
In the hybrid model  the unknown strong phases prohibit a meaningful calculation of $A_7/A_7^\prime$ in the SM \cite{Fajfer:1998dv}. We therefore do not employ constraints from eq.~(\ref{eq:A7prime}) in this model.

We consider now the impact of the chromomagnetic dipole operator.
The CP asymmetry induced by  the matrix element of  $Q_8^{(\prime)}$ estimated  within LCSR  reads \cite{Lyon:2012fk}
\begin{align}\label{eq:ACP_Q8p}
 A_{CP}\big|_{\langle Q_8,Q_8'\rangle}\sim-\mathrm{Im}\left[2C_8+\frac12C_8'\right] \, , 
\end{align}
which, together with data eq.~(\ref{eq:ACP}), yields the constraint $|\mathrm{Im}[\delta C_8^{(\prime)}]|\lesssim\mathcal O(0.1)$.
CP violation in charm  is also constrained by data on $\Delta A_{CP}=A_{CP}(K^+K^-)-A_{CP}(\pi^+\pi^-) \sim-2\mathrm{Im}[C_8-C_8']\sin\delta_{KK-\pi\pi}$ \cite{Giudice:2012qq,Lyon:2012fk}, where
$\Delta A_{CP}=-0.00134\pm0.00070$ \cite{Amhis:2016xyh}. 
To escape a potentially strong bound on ${\rm Im}[\delta C_8- \delta C_8']$ at permille level requires suppression by the unknown   strong phase difference between $K^+K^-$ and $\pi^+\pi^-$, $\delta_{KK-\pi\pi}$,
or some cancellations between different sources of BSM CP violation.

The Wilson coefficients $\delta C_7^{(\prime)}(M)$ and $\delta C_8^{(\prime)}(M)$ are generically related within BSM models, where $M$ denotes the matching scale, of the order
of the electroweak scale or higher.
In addition,  $\delta C_{7,8}^{(\prime)}$ are related by the renormalization group evolution.
At one-loop QCD
\begin{align}
 &\delta C_7^{(\prime)}(\mu_c)=a_7\,\delta C_7^{(\prime)}(M)+\frac{16}3(a_7-a_8)\delta C_8^{(\prime)}(M)\,,&&\delta C_8^{(\prime)}(\mu_c)=a_8\,\delta C_8^{(\prime)}(M)\,,
\end{align}
where
\begin{align}
 &a_7=\left(\frac{\alpha_s(M)}{\alpha_s(\mu_t)}\right)^{16/21}\left(\frac{\alpha_s(\mu_t)}{\alpha_s(\mu_b)}\right)^{16/23}\left(\frac{\alpha_s(\mu_b)}{\alpha_s(\mu_c)}\right)^{16/25}\,,\nonumber\\
 &a_8=\left(\frac{\alpha_s(M)}{\alpha_s(\mu_t)}\right)^{14/21}\left(\frac{\alpha_s(\mu_t)}{\alpha_s(\mu_b)}\right)^{14/23}\left(\frac{\alpha_s(\mu_b)}{\alpha_s(\mu_c)}\right)^{14/25}\,.\label{eq:a7a8}
\end{align}
Including effects of $\langle Q_8^{(\prime)}\rangle$, eq.~(\ref{eq:C7_Q8_eff}), and neglecting  SM contributions, we find
\begin{align}
 \delta A_7^{(\prime)}\big|_{Q_8^{(\prime)}}&\simeq0.4\,\delta C_7^{(\prime)}(1\,\text{TeV})-(0.3+0.1i)\,\delta C_8^{(\prime)}(1\,\text{TeV})\,.
\end{align}
Additionally, we find the mixing via $Q_{3-6}^{(\prime)}$, \cite{deBoer:2016dcg} and appendix \ref{app:C7eff}, neglecting the SM,
\begin{align}
 \delta A_7^{(\prime)}\big|_{Q_{3-6}^{(\prime)}}&\simeq(0.3-0.1i)\,\delta C_3^{(\prime)}(1\,\text{TeV})+(0.7+0.1i)\,\delta C_4^{(\prime)}(1\,\text{TeV})\nonumber\\
 &+(-3.5-1.9i)\,\delta C_5^{(\prime)}(1\,\text{TeV})+(-0.6+1.1i)\,\delta C_6^{(\prime)}(1\,\text{TeV})\,.
\end{align}
BSM effects   from  4-quark operators are, however, strongly
constrained by $\epsilon^\prime/\epsilon$ and $D-\bar D$ mixing, and we do not consider this possibility any further.

To compete with the  SM  $\delta C^{(\prime)}(M) \sim {\cal{O}}(0.1-1)$ is required, which is difficult to achieve given the loop factor
and possible  further flavor suppressions. However,
BSM CP asymmetries around a percent  require   $\delta C(M) $ of a few permille only but need sizable phases.
The impact of  $\delta C^\prime(M) $ on   CP asymmetries is suppressed due to the hierarchy between the left and right-chiral  SM contribution and since there is no interference between them in the branching ratio.

\subsection{Leptoquark models}\label{sec:leptoquarks}

We consider contributions from scalar $S_{1,2,3}$ and vector $V_{1,2,3}, \tilde V_{1,2}$  leptoquark representations to $c \to u \gamma$ processes, see \cite{Buchmuller:1986zs,Davidson:1993qk,deBoer:2015boa,Dorsner:2016wpm,Hiller:2016kry} for  Lagrangians and details~\footnote{In \cite{deBoer:2015boa} the notation differs from the one used here by means of charge conjugated fields. Here we write $q\to\bar q^C$ for the leptoquarks $S_1$, $S_3$, $V_2$ and $\tilde V_2$ in \cite{deBoer:2015boa} and adjust their couplings
correspondingly. Moreover, here an additional sign for all vector leptoquarks  is accounted for. Conclusions in \cite{deBoer:2015boa} are unaffected.}.
 In this section we denote  by $M$ the mass of the leptoquark and 
by $\lambda_{L/R}$  leptoquark couplings to left-/right-handed leptons.
To simplify the notation this includes also neutrinos with appropriate replacements.
For  vector-like couplings we omit the chirality index. 

A matching at $\mu\sim M$ yields, employing expressions from \cite{Lavoura:2003xp},
$\delta_SC_7^{(\prime)}(\mu=M)=1/(2\sqrt2G_F)([-Q_l/12-Q_S/24]\kappa^{(\prime)}/M^2+m_l/m_c[-Q_S/4+Q_l(3/4+1/2\ln[\mu_M^2/M^2])]\nu^{(\prime)}/M^2)$ and $\delta_VC_7^{(\prime)}(\mu=M)=1/(2\sqrt2G_F)[Q_l/3+Q_V5/12]\kappa^{(\prime)}/M^2$,
where $Q_l$ denotes the electric charge of the lepton.
The electric charge of the leptoquark $Q_{S,V}$ is fixed by charge conservation and the following vector $(V)$  and scalar $(S)$ operators are induced at tree-level 
\begin{align}
O_V^{(l)}=(\bar u_L\gamma_{\mu}l_L)(\overline l_L\gamma^{\mu}c_L) \, , \quad O_S^{(l)}=(\bar u_Ll_R)(\overline l_Lc_R)
\end{align}
plus  chirality-flipped contributions.
Here, $C_V^{(l)(\prime)}(\mu =M)=\kappa^{(\prime)}/M^2$ and $C_S^{(l)(\prime)}(\mu=M)=\nu^{(\prime)}/M^2$.
The couplings $\kappa^{(\prime)}$ and $\nu^{(\prime)}$ within leptoquark models are given in table~\ref{tab:LQ_C7prime_couplings}.
At one-loop QCD
$C_V^{(l)(\prime)}(\mu_c)=C_V^{(l)(\prime)}(M)$, $C_S^{(l)(\prime)}(\mu)=(\alpha_s(M)/\alpha_s(\mu))^{-4\beta_0}\,C_S^{(l)(\prime)}(M)$ and $\delta C_7^{(\prime)}(\mu)=(\alpha_s(M)/\alpha_s(\mu))^{16/(3\beta_0)}\delta C_7^{(\prime)}(M)-3/56[(\alpha_s(M)/\alpha_s(\mu))^{16/(3\beta_0)}-(\alpha_s(M)/\alpha_s(\mu))^{-4/\beta_0}]C_S^{(l)(\prime)}(M)/(2\sqrt2G_F)$, 
where $\beta_0=11-2/3\,n_f$ and $n_f$ is the number of active flavors, hence thresholds need to be taken into account.

\begin{figure}[!htb]
 \centering
 \includegraphics[trim=0cm 23.5cm 9cm 3.5cm,clip,width=0.8\textwidth]{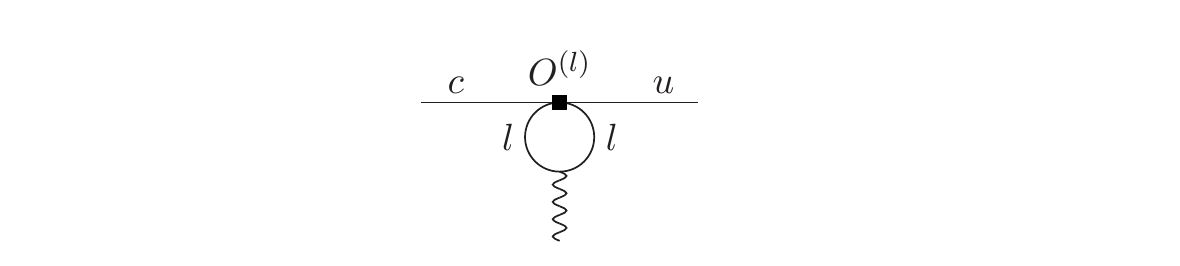}
 \caption{Diagram inducing $c\to u\gamma$ within leptoquark models.} 
 \label{fig:diagram_LQ}
\end{figure}

At the scale $\mu=m_\tau$ the $\tau$ lepton is to be integrated out. Since numerically $m_\tau\sim \sqrt2m_c$ we include  the tau-loop contributions in the 
matrix element of $O_{V,S}^{(l)}$, see figure~\ref{fig:diagram_LQ}.
The contribution of  $O_V^{(l)(\prime)}$ vanishes to all orders in $\alpha_s$. 
Including the matrix element of  $O_S^{(l)(\prime)}$ we obtain
\begin{align}\label{eq:LQ_A7prime_RL}
 \delta_SA_7^{(\prime)}(\mu_c)&=\frac1{2\sqrt2G_F}\bigg[\left(\frac{\kappa^{(\prime)}}{M^2}\left(-\frac{Q_S}{24}-\frac{Q_l}{12}\right)+\frac{\nu^{(\prime)}}{M^2}\frac{m_l}{m_c}\left(-\frac{Q_S}4+Q_l\left(\frac{45}{56}+\frac12\ln\frac{\mu_M^2}{M^2}\right)\right)\right)a_7\nonumber\\
 &+\frac{\nu^{(\prime)}}{M^2}\frac{m_l}{m_c}Q_l\left(-\frac3{56}-\frac12\ln\frac{\mu_c^2}{m_l^2}\right)a_S\bigg]\,,\\
 \delta_VA_7^{(\prime)}(\mu_c)&=\frac1{2\sqrt2G_F}\frac{\kappa^{(\prime)}}{M^2}\left(\frac{5Q_V}{12}+\frac{Q_l}3\right)a_7\,.
\end{align}
Here,
\begin{equation}
 a_S=\left(\frac{\alpha_s(M)}{\alpha_s(\mu_t)}\right)^{-12/21}\left(\frac{\alpha_s(\mu_t)}{\alpha_s(\mu_b)}\right)^{-12/23}\left(\frac{\alpha_s(\mu_b)}{\alpha_s(\mu_c)}\right)^{-12/25}
\end{equation}
and $a_7$ is given in eq.~(\ref{eq:a7a8}).
The resulting coefficients are worked out numerically in table \ref{tab:LQ_A7prime}.
Finite contributions are expected at two-loop QED.
The corresponding  leading order calculation in $\alpha_e$ is similar to the $\langle Q_2\rangle$ two-loop QCD calculation in \cite{Greub:1996wn,Greub:1996tg}, and beyond the scope of our work.
Note that $\delta_{LQ}A_8^{(\prime)}(\mu_c)$ is additionally $\alpha_e/(4 \pi)$ suppressed and will be neglected throughout.
\begin{table}[!htb]
 \centering
 \begin{tabular}{c|c|c|c|c}
  LQ            &  $\kappa$                                &  $\kappa'$                                &  $\nu$                                   &  $\nu'$  \\
  \noalign{\hrule height 1pt}
  $S_1$         &  $\lambda_R^{(cl)}(\lambda_R^{(ul)})^*$  &  $\lambda_L^{(cl)}(\lambda_L^{(ul)})^*$   &  $\lambda_L^{(cl)}(\lambda_R^{(ul)})^*$  &  $\lambda_R^{(cl)}(\lambda_L^{(ul)})^*$  \\
  $S_2$         &  $(\lambda_R^{(cl)})^*\lambda_R^{(ul)}$  &  $(\lambda_L^{(cl)})^*\lambda_L^{(ul)}$   &  $(\lambda_L^{(cl)})^*\lambda_R^{(ul)}$  &  $(\lambda_R^{(cl)})^*\lambda_L^{(ul)}$  \\
  $S_3$         &  --                                      &  $\lambda^{(cl)}(\lambda^{(ul)})^*$       &  --                                      &  --  \\
  \hline
  $V_1$         &  --                                      &  $-(\lambda^{(c\nu)})^*\lambda^{(u\nu)}$  &  --                                      &  --  \\ 
  $\tilde V_1$  &  $-(\lambda^{(cl)})^*\lambda^{(ul)}$     &  --                                       &  --                                      &  --  \\
  $V_2$         &  $\lambda^{(cl)}(\lambda^{(ul)})^*$      &  --                                       &  --                                      &  --  \\
  $\tilde V_2$  &  --                                      &  $\lambda^{(cl)}(\lambda^{(ul)})^*$       &  --                                      &  --  \\
  $V_3$         &  --                                      &  $-2(\lambda^{(cl)})^*\lambda^{(ul)}$     &  --                                      &  --  \\
 \end{tabular}
 \caption{Couplings to eq.~(\ref{eq:LQ_A7prime_RL}) with flavor indices made explicit.
 The lepton index $l$ also represents neutrinos in the models $S_2$ (for $\kappa'$), $S_3$ (with an additional factor 2), $\tilde V_2$ and $V_3$.}
 \label{tab:LQ_C7prime_couplings}
\end{table}

\begin{table}[!htb]
 \centering
 \begin{tabular}{c|c|c}
  LQ            &  $\delta_{LQ}A_7$                                                                                            &  $\delta_{LQ}A_7'$  \\
  \noalign{\hrule height 1pt}
  $S_1$         &  $-0.001\,\lambda_R^{(cl)}(\lambda_R^{(ul)})^*+0.04\tfrac{m_l}{m_\tau}\lambda_L^{(cl)}(\lambda_R^{(ul)})^*$  &  $-0.001\,\lambda_L^{(cl)}(\lambda_L^{(ul)})^*+0.04\tfrac{m_l}{m_\tau}\lambda_R^{(cl)}(\lambda_L^{(ul)})^*$  \\
  $S_2$         &  $0.002\,(\lambda_R^{(cl)})^*\lambda_R^{(ul)}-0.03\tfrac{m_l}{m_\tau}(\lambda_L^{(cl)})^*\lambda_R^{(ul)}$   &  $0.004\,(\lambda_L^{(cl)})^*\lambda_L^{(ul)}-0.03\tfrac{m_l}{m_\tau}(\lambda_R^{(cl)})^*\lambda_L^{(ul)}$  \\
  $S_3$         &  0                                                                                                           &  $-0.003\,\lambda^{(cl)}(\lambda^{(ul)})^*$  \\
  \hline
  $V_1$         &  0                                                                                                           &  $0.003\,(\lambda^{(c\nu)})^*\lambda^{(u\nu)}$  \\
  $\tilde V_1$  &  $0.01\,(\lambda^{(cl)})^*\lambda^{(ul)}$                                                                    &  0  \\
  $V_2$         &  $0.005\,(\lambda^{(cl)})^*\lambda^{(ul)}$                                                                   &  0  \\
  $\tilde V_2$  &  0                                                                                                           &  $0.01\,\lambda^{(cl)}(\lambda^{(ul)})^*$  \\
  $V_3$         &  0                                                                                                           &  $0.04(\lambda^{(cl)})^*\lambda^{(ul)}$  \\
 \end{tabular}
 \caption{Leptoquark induced coefficients for $M=1\,\text{TeV}$  from eq.~(\ref{eq:LQ_A7prime_RL}).
 For $M \to 10\,\text{TeV}$ the effective coupling scales as $\lambda^{(cl)}\lambda^{(ul)}/(M=1\,\text{TeV})^2\to0.9\,\lambda^{(cl)}\lambda^{(ul)}/(M=10\,\text{TeV})^2$.
 The lepton index $l$ also represents neutrinos in the models $S_{2,3}$, $\tilde V_2$ and $V_3$.}
 \label{tab:LQ_A7prime}
\end{table}

Constraints on $\tau$ couplings are worked out and given in table~\ref{tab:LQ_tau_constraints}, where we followed \cite{deBoer:2015boa} and used  \cite{Olive:2016xmw}.
Note that $\mathcal B(D^0\to\rho^0\gamma)$ yields no constraint for $\lambda\lesssim1$.
Lepton flavor violating $\tau$ decays constrain couplings with a $\tau$ and a light lepton; we do not take these constraints into account as they can be evaded with a flavor suppression
of the light leptons.

Numerically, the largest contributions are found for  $l=\tau$ and the constraints given  in table \ref{tab:LQ_tau_constraints}, and read
\begin{align} \label{eq:resum}
 |\delta_SA_7^{(\prime)}(\mu_c)|\simeq0.04|\lambda_R\lambda_L^*| (\mbox{TeV}/M)^2 \lesssim  {\cal{O}}(10^{-2}) \, ,
\end{align}
for the chirality-flipping contributions $\propto m_\tau/m_c$ of the leptoquark representations $S_1,S_2$.
As we are  interested in CP asymmetries we allow here for a mild suppression of the real parts of $\lambda_R \lambda_L^*$ relative to the imaginary ones, the latter of which are weaker constrained experimentally.

\begin{table}[!htb]
 \centering
 \begin{tabular}{c|c|c}
  couplings/mass                                                               &  constraint                &  observable  \\
  \noalign{\hrule height 1pt}
  $|\lambda_{S_3}^{(u\tau)}|$                                                  &  $\sim[0.0,0.2]$           &  $\tau^-\to\pi^-\nu_\tau$  \\
  $\mathrm{Re}[\lambda_{SR}^{(u\tau)}(\lambda_{SL}^{(u\tau)})^*]$              &  $\sim[0.00,0.09]$         &  \\
  $|\lambda_{V_{1,3}}^{(u\tau)}|$                                              &  $\sim[0.0,0.4]$           &  \\
  \hline
  $\mathrm{Re}[\lambda_{S_1L,S_3}^{(u\tau)}(\lambda_{S_1L,S_3}^{(c\tau)})^*]$  &  $\sim[-0.2,0.2]$          &  $\tau^-\to K^-\nu_\tau$  \\
  $\mathrm{Re}[\lambda_{SR}^{(u\tau)}(\lambda_{SL}^{(c\tau)})^*]$              &  $\sim[-0.07,0.04]$        &  \\
  $|\mathrm{Im}[\lambda_{SR}^{(u\tau)}(\lambda_{SL}^{(c\tau)})^*]|$            &  $\sim[0.0,0.7]$           &  \\
  $\mathrm{Re}[\lambda_{V_{1,3}}^{(u\tau)}(\lambda_{V_{1,3}}^{(c\tau)})^*]$    &  $\sim[-0.1,0.1]$          &  \\
  $|\mathrm{Im}[\lambda_{V_{1,3}}^{(u\tau)}(\lambda_{V_{1,3}}^{(c\tau)})^*]|$  &  $\sim[0.0,0.9]$           &  \\
  \hline
  $|\mathrm{Re}[\lambda_{SL,SR}^{(u\tau)}(\lambda_{SL,SR}^{(c\tau)})^*]|$      &  $\sim[0,0.02]$            &  $\Delta m_{D^0}$  \\
  $|\mathrm{Re}[\lambda_{S_3}^{(u\tau)}(\lambda_{S_3}^{(c\tau)})^*]|$          &  $\sim[0,0.007]$           &  \\
  $|\mathrm{Re}[\lambda_{V_2}^{(u\tau)}(\lambda_{V_2}^{(c\tau)})^*]|$          &  $\sim[0,0.8]$             &  \\
  $|\mathrm{Re}[\lambda_{V_3}^{(u\tau)}(\lambda_{V_3}^{(c\tau)})^*]|$          &  $\sim[0,0.4]$             &  \\
  \hline
  $\mathrm{Re}[\lambda_{SR}^{(u\tau)}(\lambda_{SL}^{(c\tau)})^*]$              &  $\lesssim0.3$             &  $D^+\to\tau^+\nu_\tau$  \\
  $|\mathrm{Re}[\lambda_{V_{1,3}}^{(u\tau)}(\lambda_{V_{1,3}}^{(c\tau)})^*]|$  &  $\lesssim0.5$             &  \\
  \hline
  $\mathrm{Re}[\lambda_{SR}^{(c\tau)}(\lambda_{SL}^{(c\tau)})^*]$              &  $\sim[-1,0.09]$           &  $D_s\to\tau^+\nu_\tau$  \\
  $|\lambda_{S_3}^{(c\tau)}|$                                                  &  $\sim[0.0,0.4]$           &  \\
  $|\lambda_{V_{1,3}}^{(c\tau)}|$                                              &  $\sim[0.0,0.2]$           &  \\
  \hline
  $|\lambda_{S_1L,S_3}^{(u\tau)}\lambda_{S_1L,S_3}^{(c\tau)}|$                 &  $\lesssim4\cdot10^{-4}$   &  $(K^+\to\pi^+\bar\nu\nu)/(K^+\to\pi^0\bar e \nu)$  \\
  $|\lambda_{V_3}^{(u\tau)}(\lambda_{V_3}^{(c\tau)})^*|$                       &  $\lesssim8\times10^{-5}$  &  \\
  \hline
 \end{tabular}
 \caption{Scalar and vector leptoquark constraints on couplings to $\tau$'s scaling as $\text{TeV}/M$ and $\sqrt{\text{TeV}/M}$ for $\Delta m_{D^0}$.
 The kaon  constraints  are found via \cite{Carpentier:2010ue}.
 The vector ($\tilde V_{1,2}$) leptoquark couplings are unconstrained by the above  observables.}
 \label{tab:LQ_tau_constraints}
\end{table}

For $l=e,\mu$ we find with the constraints given in \cite{deBoer:2015boa} that $|\delta_{\tilde V_1}A_7| _{e,\mu}\lesssim0.0006$, $|\delta_{V_2}A_7|_{e,\mu}\lesssim0.0003$, $|\delta_{\tilde V_2}A_7'|_{e,\mu}\lesssim0.0006$ and $|\delta_{V_3}A_7'|_{e,\mu}\lesssim0.003$.
The contributions from scalar leptoquarks are negligible, $|\delta_{S_{1,2,3}}A_7^{(\prime)}|_{e,\mu} \lesssim\mathcal O(10^{-5})$.

The $\tau$ couplings for $S_{1L}$, $S_3$ and $V_3$ receive their strongest constraint from $K$ decays, $|\delta_{S_{1L,3}}A_7'|_{\tau} \lesssim\mathcal O(10^{-7})$ and $|\delta_{V_3}A_7'|_{\tau} \lesssim\mathcal O(10^{-6})$.
All other  bounds in  table~\ref{tab:LQ_tau_constraints} can be escaped  with phase-tuning  ${\rm Im }[\lambda^{(u \tau)}(\lambda^{(c\tau)})^*] \gg {\rm Re} [\lambda^{(u \tau)}(\lambda^{(c\tau)})^*]$. 
Corresponding BSM coefficients $\delta A_7$ and $\delta A_7^\prime$ can be read off from table~\ref{tab:LQ_A7prime} for $ \lambda_{L,R} ^{(q \tau)} \lesssim 1$.
Assuming instead  ${\rm Im }[\lambda^{(u \tau)}(\lambda^{(c\tau)})^*]  \lesssim  {\rm Re} [\lambda^{(u \tau)}(\lambda^{(c\tau)})^*]$, one obtains from $D$-mixing $|\delta_{S_{1L,2L,2R}}A_7^{(\prime)}|_{\tau} \lesssim\mathcal O(10^{-5})$. 
For the $SU(2)$-singlet and -doublet vector leptoquarks  we find $|\delta_{V_1}A_7|_{v_\tau} \lesssim 0.0003$, $|\delta_{\tilde V_1}A_7|_{\tau} \lesssim  0.01$, $|\delta_{V_2}A_7|_{\tau} \lesssim 0.004$, and $|\delta_{\tilde V_2}A_7'|_{\tau} \lesssim 0.01$.
Note, $\delta_{S_1,S_2}  A_7^{(\prime)}|_{\tau}$ from chirality-flipping contributions without  resummation would be about one  order of magnitude larger than eq.~(\ref{eq:resum}).

To summarize, within leptoquark models the $c\to u\gamma$ branching ratios are SM-like with CP asymmetries at $\mathcal O(0.01)$   for $V_2$ and SM-like for $S_3$ and $V_{1,3}$.
On the other hand, in  models $S_{1,2}$ and $\tilde V_{1,2}$ $A_{CP} \lesssim\mathcal O(10 \%)$.
\footnote{For vector leptoquarks, which are gauge bosons in a renormalizable model,  the coupling matrix is  unitary, see  {\it e.g.} \cite{Davidson:1993qk}.  
In this case, $\delta_V C_{7,8}^{(\prime)}(M) =0$ as a result of a GIM-like mechanism. The argument could in principle be invalidated if there would be contributions from both lepton chiralities
$\propto \lambda_R \lambda_L^*$, however,  this does not happen in charm FCNCs for vector leptoquarks (we do not consider right-handed neutrinos). As a result, the corresponding CP asymmetries are SM-like.}
The largest effects arise from $\tau$-loops.

\subsection{SUSY}\label{sec:smia}

Here we consider effects within SUSY, taking into account the leading, gluino induced contributions   within the mass insertion approximation
 \cite{Gabbiani:1996hi,Prelovsek:2000xy} 
\begin{align}
 &\delta_\text{SUSY}C_7^{(\prime)}(M)=-\frac{16}{9\sqrt2G_F}\frac{\alpha_s\pi}{m_{\tilde q}^2}  (\delta_{12})_{LR(RL)}\,\frac{m_{\tilde g}}{m_c} M_1(x) \,,\nonumber\\
 &\delta_\text{SUSY}C_8^{(\prime)}(M)=-\frac1{\sqrt2G_F}\frac{\alpha_s\pi}{m_{\tilde q}^2}   (\delta_{12})_{LR(RL)}\,\frac{m_{\tilde g}}{m_c}\left(-\frac13M_1(x) -3M_2(x)\right)   \,,
\end{align}
where $m_{\tilde q}$ and $m_{\tilde g}$ denote the masses of the squarks and the gluino, respectively,  $M \sim  m_{\tilde g, \tilde q}$,
\begin{align}
 &M_1(x)= \frac{1+4x-5x^2+4x\ln x+2x^2\ln x}{2(1-x)^4}\,,&&M_1(x=1)=\frac1{12}   \,,\nonumber\\
&M_2(x)=-x^2 \frac{5-4x-x^2+2\ln x+4x\ln x}{2(1-x)^4}\,,&&M_2(x=1)= \frac1{12}\,,
\end{align}
and $x=m_{\tilde g}^2/m_{\tilde q}^2$. We neglected terms not subject to $m_{\tilde g}/m_c$-enhancement.

The mass insertions $(\delta_{12})$ are constrained by data on the $D^0-\bar D^0$ mass difference  \cite{Gabbiani:1996hi,Olive:2016xmw}. In case of 
{\it i)}: $m_{\tilde g}^{(i)}=1\,\text{TeV},\,m_{\tilde q}^{(i)}=2\,\text{TeV}$ and {\it ii)}: $m_{\tilde g}^{(ii)}=2\,\text{TeV},\,m_{\tilde q}^{(ii)}=1\,\text{TeV}$, we find 
\begin{align}
  \sqrt{\left|\mathrm{Re}(\delta_{12}^{(i)})_{LR,RL}^2\right|}\lesssim0.07 \, , \quad \quad 
  \sqrt{\left|\mathrm{Re}(\delta_{12}^{(ii)})_{LR,RL}^2\right|}\lesssim0.02\,,
\end{align}
and $|\delta_\text{SUSY}^{(i)}A_7^{(\prime)}|\lesssim0.2$ and $|\delta_\text{SUSY}^{(ii)}A_7^{(\prime)}|\lesssim0.3$, respectively.
The upper limits are similar to  the model-independent constraints obtained in section \ref{sec:model_independently}. 
Note that, barring cancellations, constraints on the imaginary parts of $\delta_{12}$ can be about an order of magnitude stronger \cite{Carrasco:2014uya}, still permitting to signal  BSM CP-violation.
Constraints from $\epsilon'/\epsilon$ and chargino loops are model-dependent.
For realistic, not too low SUSY mass parameters corresponding bounds, {\it e.g.} \cite{Frank:2003ra}, are not stronger than those from $D-\bar D$ mixing.
We further checked that $|\delta_\text{SUSY}^{(i),(ii)}C_{3-6}^{(\prime)}(M)|\lesssim10^{-5}$ is negligible.
Note that $|\delta_\text{SUSY}A_7^{(\prime)}/\delta_\text{SUSY}A_8^{(\prime)}|\in[0.75,5.46]$, where the lower limit is  for $x\gtrsim10$ and the upper limit for $x\lesssim0.01$.
Thus, supersymmetric models may induce $c\to u\gamma$ branching ratios and CP asymmetries above the SM predictions, and SUSY parameters are constrained by 
 $\mathcal B(D^0\to\rho^0\gamma)$.
Note that  additional constraints may apply once the SUSY breaking has been specified  \cite{Prelovsek:2000xy}.
A detailed evaluation is beyond the scope of this work.

\section{On \texorpdfstring{$\Lambda_c\to p\gamma$}{Lambdactopgamma} \label{sec:pol}}

We investigate possibilities  to probe  the handedness of the $c\to u\gamma$ current   in  the decay $\Lambda_c\to p\gamma$ with polarization asymmetries, that arise once 
$\Lambda_c$'s are produced polarized.  
 We follow closely related  works on $\Lambda_b \to \Lambda \gamma$ decays \cite{Hiller:2001zj,Falk:1993rf}.

The $\Lambda_c \to p \gamma$ branching ratio is not measured to date. Quite generally we assume
\begin{align}\label{eq:B_Lambdacpgamma}
 \mathcal B (\Lambda_c\to p\gamma) \sim\mathcal O(10^{-5})\,,
\end{align}
in agreement with  naive expectations from 
$\mathcal B(D^0\to\rho^0\gamma)$
\footnote{Via  weak annihilation $r(\Lambda_c\to p\gamma/D^0\to\rho^0\gamma)\sim\sqrt2$ due to color counting, via resonances $r(\Lambda_c\to p\gamma/D^0\to\rho^0\gamma)\sim1$ due to the amplitude $A^{\text{III},\Lambda_c\to p\gamma}$ \cite{Singer:1996ba} and $r(\Lambda_c\to p\gamma/D^0\to\rho^0\gamma)\simeq\sqrt2f_\perp/T\sim1$ via SM effective and BSM Wilson coefficients.
Here, the form factor $f_\perp=f_\perp^T(0)=f_\perp^{T5}(0)$ is defined as in \cite{Boer:2014kda} and calculated within  QCD LCSR
\cite{Khodjamirian:2011jp,Li:2016qai}, a covariant confined quark model \cite{Gutsche:2014zna} and a relativistic quark model \cite{Faustov:2016yza}.
Within a constituent quark model $\mathcal B_{\Lambda_c\to p\gamma}=2.2\cdot10^{-5}$   \cite{Uppal:1992cc} in agreement with eq.~(\ref{eq:B_Lambdacpgamma}).}.
Note, we employ equation (\ref{eq:B_Lambdacpgamma}) only to estimate uncertainties. ${\cal{B}}(\Lambda_c\to p\gamma)$ should be determined
experimentally.

 The number of $\Lambda_c\to p\gamma$  events $N$, modulo reconstruction efficiencies, can be obtained from
 \begin{align} 
 N=N(c\bar c)\,f(c\to\Lambda_c)\,\mathcal B (\Lambda_c\to p\gamma) \sim N(c\bar c)\cdot10^{-6}\,,
\end{align}
where $f(c\to\Lambda_c)\simeq0.06$ \cite{Lisovyi:2015uqa} is the fragmentation fraction of charm to $\Lambda_c$-baryons and $N(c\bar c)$  the number of $c\bar c$ produced.
At the forthcoming Belle II experiment, where  $\sigma(e^+e^-\to\bar cc)\simeq1.3\,\text{nb}$, $L\simeq5\,\text{ab}^{-1}$ within a year \cite{Aushev:2010bq}, 
$N \sim [10^3 ,10^4]$.
At a future $e^+ e^-$-collider running at the $Z$ (FCC-ee), where $N(Z)\sim10^{12}$ within one year  \cite{dEnterria:2016fpc} and $\mathcal B(Z\to c\bar c)\simeq0.12$ \cite{Olive:2016xmw}, $N\sim10^5$. This environment suggests a measurement of the $\Lambda_c \to p \gamma$ branching ratio, the $\Lambda_c$ polarization and the angular asymmetry $A^\gamma$  of $\Lambda_c\to p\gamma$ decays.
The latter is defined in the $\Lambda_c$ rest frame by the angle between the $\Lambda_c$ spin and the proton momentum, that is, the forward-backward asymmetry of the photon momentum relative to the $\Lambda_c$ boost, and normalized to the width. It reads \cite{Hiller:2001zj}
\begin{align}\label{eq:Agamma}
 A^\gamma=-\frac{P_{\Lambda_c}}2\frac{1-|r|^2}{1+|r|^2}\,,
\end{align}
where $P_{\Lambda_c}$ denotes  the (longitudinal) $\Lambda_c$ polarization and $r=A_7'/A_7$.  $ A^\gamma \to  P_{\Lambda_c}/2$ for  $r\to\infty$ and
 $ A^\gamma \to  -P_{\Lambda_c}/2$ for  $r\to0$. Calculating $A_7, A_7^\prime$ in the SM is a difficult task
and beyond the scope of our work. In the  subsequent estimates of BSM sensitivity we assume that approximately $ A_7^{(\prime)} \sim \delta C_7^{(\prime)}$ for large BSM effects and  $A_7^\prime \ll A_7$ in SM-like situations.
$A^\gamma$ is measurable in the laboratory frame for a boost $\vec\beta=\vec p_{\Lambda_c}/E_{\Lambda_c}$, where $\vec p_{\Lambda_c}$ denotes the $\Lambda_c$ three-momentum in the laboratory frame, as 
\begin{align}
 \langle q_\parallel\rangle_{|\vec\beta|}=\gamma E_\gamma^*\left(\left|\vec\beta\right|+\frac23A^\gamma\right) \, .
\end{align}
Here, $\langle q_\parallel\rangle_{|\vec\beta|}$ is the average longitudinal momentum of the photon in the laboratory frame relative to the boost axis, $\gamma=1/\sqrt{1-|\vec\beta|^2}$ and $E_\gamma^*=(m_{\Lambda_c}^2-m_p^2)/(2m_{\Lambda_c})\simeq0.95\,\text{GeV}$ is the photon energy in the $\Lambda_c$ rest frame.

The $\Lambda_c$ polarization can be expressed in terms of the charm quarks' polarization $P_c$ as \cite{Falk:1993rf,Galanti:2015pqa,Kats:2015zth}
\begin{align}\label{eq:PLambdacPc}
 \frac{P_{\Lambda_c}}{P_c}\simeq\frac{1+A(0.07+0.46\omega_1)}{1+A}   \simeq0.68\pm0.03 \, , 
\end{align}
where $A\simeq1.1$ is extrapolated \cite{Galanti:2015pqa} from a measurement by the E791 collaboration \cite{Aitala:1996cy} and $\omega_1=0.71\pm0.13$ is measured by the CLEO collaboration \cite{Brandenburg:1996jc}.
At the $Z$ the polarization of the charm-quark $P_c^{(Z)}\simeq-0.65$ \cite{Korner:1993dy} and one obtains a sizable polarization
\begin{align} \label{eq:PZ}
 P_{\Lambda_c}^{(Z)}\simeq-0.44\pm0.02\,.
\end{align}
The polarization is negative since $c$-quarks from the $Z$ are predominantly left-handed. Ultimately, its value needs to be determined experimentally,
 {\it e.g.,} from   $\Lambda_c\to\Lambda(\to p\pi)l\nu$ decays with $\mathcal B(\Lambda_c\to\Lambda(\to p\pi)e\nu_e)\simeq0.023$ \cite{Olive:2016xmw}.
Note that  the depolarization parameters $A$ and $\omega_1$ are measurable at Atlas, BaBar, Belle, CMS and LHCb \cite{Galanti:2015pqa}.
The $\Lambda_c$ polarization itself is measurable at Atlas, CMS and LHCb via $pp\to t(\to bW^+(\to c\bar s))\bar t(\to\bar bW^-(\to l^-\bar\nu))$ \cite{Galanti:2015pqa} and $pp\to W^-c$ \cite{Kats:2015zth}, where $P_c^{(W^-)}\simeq-0.97$ \cite{Korner:1993dy}.

\begin{figure}[!htb]
 \centering
 \includegraphics[width=0.8\textwidth]{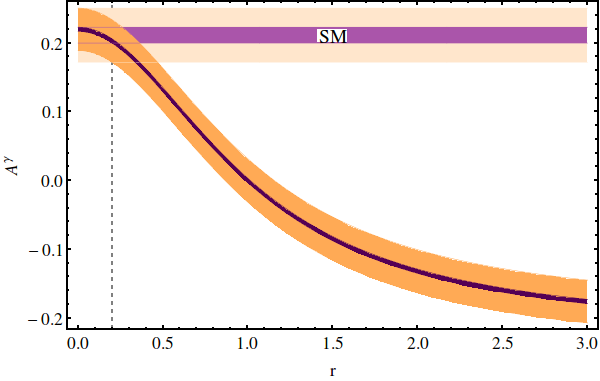}
 \caption{The angular asymmetry (\ref{eq:Agamma}) of $\Lambda_c\to p\gamma$  decays as a function of $r=A_7'/A_7$ for $P_{\Lambda_c}=-0.44$.
 The bands represent the statistical uncertainties for $N=10^3$ (orange) and $N=10^5$ (purple).
 Within the SM and leptoquark models $r\lesssim0.2$, indicated by the dashed vertical line, and corresponding $A^\gamma$-ranges shown by the horizontal bands. Within SUSY all values $r \lesssim\mathcal O(1)$ are possible.}
 \label{fig:AgammaZ}
\end{figure}

The angular asymmetry is shown in figure~\ref{fig:AgammaZ} for $P_{\Lambda_c}$ as expected at the $Z$, eq.~(\ref{eq:PZ}). In the SM and leptoquark models
$r\lesssim0.2$, and $A^\gamma \sim -P_{\Lambda_c}/2$ and positive.
The statistical uncertainty $\delta A^\gamma=\sqrt{1-(A^\gamma)^2}/\sqrt N$ is represented by the bands corresponding to $N=10^3$ (orange) and $N=10^5$ (purple).
Within SUSY already for $N=10^3$ the angular asymmetry could be observed  essentially everywhere within $|A^\gamma| \lesssim |P_{\Lambda_c}|/2$,
allowing to signal BSM physics.
Prerequisite for an interpretation is an experimental determination  of  $P_{\Lambda_c}$ or the depolarization fraction, if $P_c$ is known.

Given the uncertainties present in meson decays our analysis of $\Lambda_c$'s is clearly explorative, pointing out an opportunity with future polarization measurements with baryons. More work is needed to  detail wrong-chirality contributions in the SM.

\section{Summary \label{sec:con}}

We worked out SM predictions for various $D_{(s)} \to V \gamma$ branching ratios, which are compiled in tables  \ref{tab:DVgamma_branching_ratios} and \ref{tab:Dphigamma_DKstar0gamma_branching_ratios}.  
The  hybrid model predicts values up to  a factor $\sim 2-3$ larger than  the QCD factorization based approach, the latter being dominated by contributions from weak annihilation. All three branching ratios  measured so far, the ones of  $D^0 \to \rho^0 \gamma, D^0 \to \phi \gamma$ and $D^0 \to \bar K^{*0} \gamma$
are above the QCD factorization range given, suggesting, to the order we are working,
a low value of the parameter $\lambda_D \lesssim 0.1$ GeV or low charm mass scale.
One has to keep in mind, however, that poor convergence of the $1/m_c$ and $\alpha_s$-expansion prohibits a sharp  conclusion without further study.
Decays of charged mesons with color allowed weak annihilation contribution are better suited for extracting $\lambda_D$ as there is lesser chance for large cancellations, see
also  figure~\ref{fig:BlambdaDDrhogamma}.
The measured branching ratios are close to the top end of the ones  obtained in the hybrid model.
$D^0 \to \phi \gamma$ and $D^0 \to \bar K^{*0} \gamma$ belong to the class of those  decays with no  direct contribution from electromagnetic  dipole operators. 
Corresponding decays  are listed in table \ref{tab:Dphigamma_DKstar0gamma_branching_ratios}, their branching ratios have essentially no sensitivity to BSM physics
unlike the CP asymmetry in $D^0\to\phi\gamma$, {\it cf.} equations (\ref{eq:ACPphi}), (\ref{eq:WACP}).

The measured $D^0 \to \rho^0 \gamma$ branching ratio provides a  model-independent upper limit on the decay amplitudes given in eq.~(\ref{eq:limit}), which is  similar to the
one  from $D \to \pi \mu \mu$ decays \cite{deBoer:2015boa}. If ${\cal{B}}(D^0 \to \rho^0 \gamma)$ is saturated with BSM physics or in the SM,
${\cal{B}}(D^0 \to (\rho^0/\omega)  \gamma)$ are very close to each other. 
For intermediate scenarios the two branching ratios can differ by orders of magnitude \cite{Fajfer:2000zx}, and indicate BSM physics.

 CP asymmetries in $c \to u \gamma$ transitions constitute SM null tests. We find  $A_{CP}^{\rm SM}  \lesssim  {\rm few} \cdot 10^{-3}$ for $D^0 \to \rho^0 \gamma$, see figure~\ref{fig:ACPB},
 and similar for other radiative rare charm decays. Among the modes in tables  \ref{tab:DVgamma_branching_ratios} 
$A_{CP}$ is measured only in $D^0 \to \rho^0 \gamma$ decays, eq.~(\ref{eq:ACP}),  consistent with  zero and the SM.
Uncertainties on $A_{CP}$  are presently too large to provide phenomenologically useful constraints. However,
CP-violating BSM can induce significant CP asymmetries at the level of  $\sim {\cal{O}}(0.1)$, and as demonstrated  in figure \ref{fig:C7BSM},
already a factor four  reduction of statistical uncertainty (with central values kept) shows that constraints  can be improved significantly.

We worked out implications for two  BSM models, SUSY and leptoquark ones. We find that SUSY can saturate the measured $D^0 \to \rho^0 \gamma$ branching ratio
and CP asymmetry while leptoquark models can't. In the latter CP asymmetries $\lesssim 10 \%$  are possible.
The largest  effects  stem from models with
vector leptoquarks  $\tilde V_1$, $\tilde V_2$ and scalars $S_1,S_2$ with  couplings to taus.

If  $\Lambda_c$-baryons are produced polarized, such as at the $Z$, angular asymmetries in $\Lambda_c \to p \gamma$ can
probe  chirality-flipped contributions, see figure~\ref{fig:AgammaZ}.
Within SUSY $A^\gamma$ can be  very different from its SM-value, including having its sign flipped.
Prerequisite for an  interpretation  of $A^\gamma$   is a measurement of  
the $\Lambda_c$ polarization, however,
irrespective of its precise value, in the SM $A^\gamma$ is expected to be positive at the $Z$.
The branching ratio of $\Lambda_c \to p \gamma$ may be investigated at Belle II \cite{Aushev:2010bq}, the depolarization fraction at the LHC \cite{Kats:2015zth}, and at the  FCC-ee  \cite{dEnterria:2016fpc} 
all of this including  $A^\gamma$.

We analyzed radiative rare charm decays in the SM and beyond with presently available technologies.
Both the heavy quark and the hybrid framework share qualitatively similar phenomenology. The reason is the  dominance of weak annihilation and corresponding
color and CKM factors.  A closer look exhibits numerical differences between the two frameworks for branching ratios and CP asymmetries, detailed in section \ref{sec:pheno}.
Despite the considerable uncertainties we demonstrated that existing charm data are already informative on BSM physics. Future measurements
can  improve theoretical frameworks and allow to check for patterns.
Given the unique window into  flavor in the up-sector that is provided
by  $|\Delta C|=|\Delta U|=1$ processes, further efforts are worthwhile and necessary.
Largest sources of parametric uncertainty  within the SM are the $\mu_c$-scale dependence and $\lambda_D$ in the heavy quark approach and
$a_{1,2}$ and the form factor $A_1(q^2)$ in the hybrid model.
For a BSM interpretation in scenarios with enhanced dipole operators improved  knowledge of $D \to V$ tensor form factors at $q^2=0$
is desirable.

\section*{Acknowledgements}

We thank Jernej Kamenik, Kamila Kowalska,  Stephane Monteil, Stefan Schacht and Alan Schwartz for useful discussions.
This work has been supported in part by the DFG Research Unit FOR 1873 ``Quark Flavour Physics and Effective Field Theories''.

\appendix

\section{Parameters}\label{app:parameters}

The couplings, the ($\overline{\text{MS}}$) masses and the widths are taken from the PDG \cite{Olive:2016xmw}, with the renormalization  scale running as in \cite{Chetyrkin:2000yt}.
The values of the CKM matrix elements are taken from the UTfit collaboration \cite{UTfit}.
The decay constants given by the FLAG read \cite{Aoki:2016frl}
\begin{align}
 &f_{D_s}=(0.24883\pm0.00127)\,\text{GeV}\,,&&f_D=(0.21215\pm0.00145)\,\text{GeV}\,,\nonumber\\
 &f_K=(0.1556\pm0.0004)\,\text{GeV}\,,&&f_\pi=(0.1302\pm0.0014)\,\text{GeV}
\end{align}
and \cite{Dimou:2012un,Straub:2015ica} (and references therein)
\begin{align}
 &f_\phi=(0.233\pm0.004)\,\text{GeV}\,,\nonumber\\
 &f_{K^*}=(0.204\pm0.007)\,\text{GeV}\,,&&f_{K^*}^\perp=(0.163\pm0.008)\,\text{GeV}\,,\nonumber\\
 &f_\rho=(0.213\pm0.005)\,\text{GeV}\,,\quad f_{\rho^0}^{(d)}=(0.2097\pm0.0003)\,\text{GeV}\,,&&f_\rho^\perp=(0.160\pm0.011)\,\text{GeV}\,,\nonumber\\
 &f_\omega=(0.197\pm0.008)\,\text{GeV}\,,\quad f_\omega^{(d)}=(0.2013\pm0.0008)\,\text{GeV}\,,&&f_\omega^\perp=(0.139\pm0.018)\,\text{GeV}
\end{align}
at $\mu=1\,\text{GeV}$ for the transverse decay constants.

The Gegenbauer moments read \cite{Dimou:2012un} (and references therein)
\begin{align}
 &a_1^{\rho\perp}=0\,,&&a_2^{\rho\perp}=0.14\pm0.06\,,\nonumber\\
 &a_1^{\omega\perp}=0\,,&&a_2^{\omega\perp}=0.14\pm0.12\,,\nonumber\\
 &a_1^{K^*\perp}=-0.04\pm0.03\,,&&a_2^{K^*\perp}=0.10\pm0.08
\end{align}
at $\mu=1\,\text{GeV}$, where the sign of $a_1^{K^*\perp}$ is due to ${K^*}^+=(u\bar s)$ \cite{Ball:2006eu}.

\section{\texorpdfstring{$D\to V$}{DtoV} form factors}\label{app:DV_form_factors}

We define the $D\to V$ form factors  as usual
\begin{align}
 \langle V(p_V,\epsilon^\nu)|\bar u\gamma_\mu(1-\gamma_5)c|D(p_D)\rangle&=-2\epsilon_{\mu\nu_1\nu_2\nu_3}(\epsilon^*)^{\nu_1}p_D^{\nu_2}q^{\nu_3}\frac{V(q^2)}{m_D+m_V}\nonumber\\
 &+i\left(q_\mu\frac{\epsilon^*\cdot q}{q^2}-\epsilon_\mu^*\right)(m_D+m_V)A_1(q^2)\nonumber\\
 &-i\left(q_\mu\frac{m_D^2-m_V^2}{q^2}-(p_D+p_V)_\mu\right)\frac{\epsilon^*\cdot q}{m_D+m_V}A_2(q^2)\nonumber\\
 &-iq_\mu(\epsilon^*\cdot q)\frac{2m_V}{q^2}A_0(q^2)\,,\nonumber\\
 \langle V(p_V,\epsilon^\nu)|\bar u\sigma_{\mu\nu}q^\nu(1+\gamma_5)c|D(p_D)\rangle&=-i2\epsilon_{\mu\nu_1\nu_2\nu_3}(\epsilon^*)^{\nu_1}p_D^{\nu_2}q^{\nu_3}T_1(q^2)\nonumber\\
 &-\left((p_D+p_V)_\mu(\epsilon^*\cdot q)-\epsilon_\mu^*\left(m_D^2-m_V^2\right)\right)T_2(q^2)\nonumber\\
 &+\left(q_\mu-(p_D+p_V)_\mu\frac{q^2}{m_D^2-m_V^2}\right)(\epsilon^*\cdot q)T_3(q^2)\,,
\end{align}
where $\epsilon_{0123}=1$ and $q_\mu=(p_D-p_V)_\mu$. In our analysis we need $V(0), T_1(0)$ and $A_1(q^2)$.

The $D\to\rho$ form factors have been measured by  the CLEO collaboration \cite{CLEO:2011ab} as
\begin{align} 
 &V(0)=0.84_{-0.11}^{+0.10}\,,\quad A_1(0)=0.56_{-0.03}^{+0.02}\,,\quad V(0)/A_1(0)=1.48\pm0.16\label{eq:Drho_ratios0}\,,
\end{align}
where statistical and systematic uncertainties are added in quadrature,
consistent with lattice computations  \cite{Flynn:1997ca} (and references therein)
\begin{align}
 &V(0)=1.1\pm0.2\;(0.71_{-0.12}^{+0.10})\,,\quad A_1(0)=0.65\pm0.07\;(0.60\pm0.06)\,.
\end{align}
Here, given in parenthesis are the preliminary computations of \cite{Gill:2001jp} with doubled uncertainties to account  for systematic uncertainties.
The  $D\to\omega$ form factors have been  measured by   the BESIII collaboration  \cite{Ablikim:2015gyp} as
(statistical and systematic uncertainties are added in quadrature)
\begin{align}\label{eq:Domega_ratios0}
 V(0)/A_1(0)=1.24\pm0.11\,,
\end{align}
consistent with the values for $D \to \rho$,  equation (\ref{eq:Drho_ratios0}).

The (axial-)vector form factors have been obtained from  QCD LCSR \cite{Wu:2006rd}
\begin{align}
 &V(0)=0.80 \pm 0.04 \,, &&A_1(0)=0.60 ^{+0.04}_{-0.03}\, ,   &&&& (D \to \rho)\nonumber \\
 &V(0)=0.74^{+0.04}_{-0.03} \,, &&A_1(0)=0.56 \pm 0.03 \, ,  &&&& (D \to \omega)\nonumber  \\
 &V(0)=0.77 \pm 0.05 \,, &&A_1(0)=0.59 \pm 0.04  &&&& (D_s \to K^*)
\end{align}
and
\begin{align}\label{eq:LCSR}
 &A_1^{(D\to K^*)}(0)=0.57 \pm 0.02 \, ,\quad A_1^{(D_s\to\phi)}(0)=0.57 \pm 0.05\,,
\end{align}
 where we rounded for easier comparison. 
The constituent quark model (CQM) \cite{Melikhov:2000yu} and the covariant light-front quark model (CLFQM) \cite{Verma:2011yw} provide $q^2$-shapes for the form factors.
Note that we do not employ  the form factors of  \cite{Fajfer:2005ug} within chiral theory as the $D\to\rho,\omega$ form factors at $q^2=0$ and  the $D\to\pi$ form factors differ from measurements/computations.

\begin{figure}[!htb]
 \centering
 \includegraphics[width=0.8\textwidth]{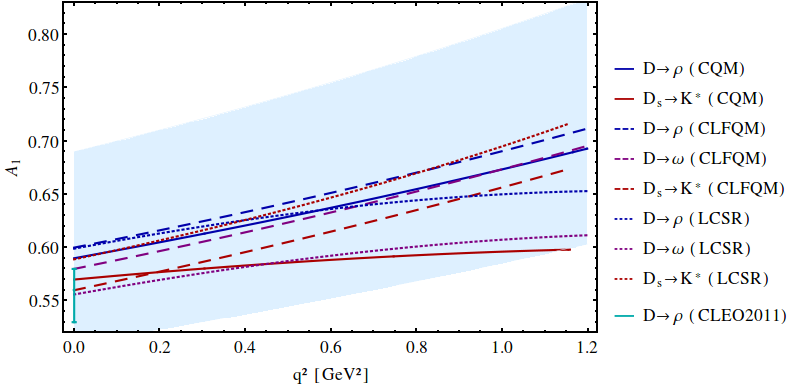}
 \\[2em]
 \includegraphics[width=0.8\textwidth]{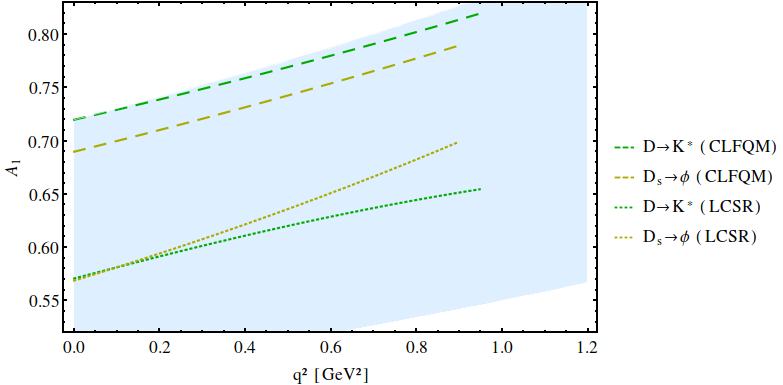}
 \caption{The  form factor $A_1(q^2)$ from the  CQM \cite{Melikhov:2000yu} (solid curves), the CLFQM  \cite{Verma:2011yw} (dashed curves) and LCSR  \cite{Wu:2006rd} (dotted curves).
 The blue, purple, red, green and yellow curves are for $D\to\rho$, $D\to\omega$, $D_s\to K^*$, $D\to K^*$ and  $D_s\to\phi$,  respectively. The curves are shown in  the 
 kinematically allowed regions only.
 The (cyan) error bar depicts the $D\to\rho$ measurement \cite{CLEO:2011ab}.
 The light blue bands represent the values employed in the numerical analysis, eq.~(\ref{eq:DV_form_factors}).}
  \label{fig:DV_formfactors}
\end{figure}

In  the numerical analysis we employ the following  form factor values
\begin{align}\label{eq:DV_form_factors}
 &T_1(0)=0.7(1\pm0.20)\,,\quad V(0)=0.9(1\pm0.25)\,,\qquad(D\to(\rho,\omega)\gamma)\nonumber\\
 &T_1(0)=0.7(1\pm0.25)\,,\quad V(0)=0.9(1\pm0.30)\,,\qquad(D_s\to K^*\gamma)\nonumber\\
 &A_1^{(c\to u)}(q^2)=\frac{0.6}{1-0.5\,q^2/m_D^2}(1\pm0.15)\,,\quad A_1^{(D\to K^*,D_s\to\phi)}(q^2)=\frac{0.6}{1-0.5\,q^2/m_D^2}(1\pm0.20)\,,
\end{align}
where uncertainties are given in parenthesis.
These ranges are consistent with eqs.~(\ref{eq:Drho_ratios0}-\ref{eq:LCSR}), the CQM and the CLFQM as well as the large energy relations  \cite{Charles:1998dr}.
$A_1(q^2)$ is shown in figure~\ref{fig:DV_formfactors}.

\section{\texorpdfstring{$D\to V\gamma$}{DtoVgamma} amplitudes in the hybrid approach}\label{app:resonant_DVgamma_amplitudes}

We write the resonance-induced contributions to  $D\to V\gamma$ amplitudes using \cite{Fajfer:1997bh,Fajfer:1998dv} and \cite{Casalbuoni:1992dx} as
\begin{align}
 A_{\text{PC}/\text{PV}}=\sqrt{\alpha_e2\pi}G_FV_{cq}^*V_{uq'}\left(A_{\text{PC}/\text{PV}}^{\text I}+A_{\text{PC}/\text{PV}}^{\text{II}}+A_{\text{PC}/\text{PV}}^{\text{III}}\right)\,.
\end{align}
 For  each $D \to V \gamma$ transition,
 the CKM factor $V_{cq}^*V_{uq'}$ 
 can be inferred from the corresponding weak annihilation contribution, eqs.~(\ref{eq:C7_WAanal}) and (\ref{eq:pureWA}). 
The amplitudes $A_{\text{PC}/\text{PV}}^{\text{III}}$ originate from the long-distance penguin  estimated with VMD. They  contain terms with different CKM factors, allowing for CP violation.
We adjusted the relative sign between the VMD contributions from $\rho, \omega$ and $\phi$ to recover $A_{\text{PC}/\text{PV}}^{\text{III}}=0$ in the $SU(3)$-limit.
For the weak annihilation modes $A_{\text{PC}/\text{PV}}^{\text{III}}=0$.

For $V=V^0\in\{\rho^0,\omega\}$ the $D^0 \to V \gamma$ amplitudes read
\begin{align}
 |A_\text{PC}^{\text I,V^0}|&=-\frac{m_{{D^*}^0}^{7/2}\left(m_{{D^*}^0}^2-(m_{D^0}+m_{\pi^0})^2\right)^{3/4}\left(m_{{D^*}^0}^2-(m_{D^0}-m_{\pi^0})^2\right)^{3/4}m_{V^0}}{\sqrt{2\pi\alpha_e}m_{D^0}^{3/2}(m_{{D^*}^0}^2-m_{D^0}^2)^{3/2}(m_{{D^*}^0}+m_{D^0}-m_{\pi^0})(m_{{D^*}^0}^2-m_{V^0}^2)}\nonumber\\
 &\times a_2f_{V^0}\sqrt{\frac{\Gamma({D^*}^0\to D^0\gamma)}{\Gamma({D^*}^0\to D^0\pi^0)}}f_+(0)\,,\nonumber\\
 |A_\text{PC}^{\text{II},V^0}|&=\frac{2\sqrt3m_{D^0}^2}{\sqrt{\alpha_e}(m_{D^0}^2-m_{\pi^0}^2)}\nonumber\\
 &\times a_2f_Df_\pi\left(\sqrt{\Gamma(\rho^0\to\pi^0\gamma)}\frac{m_\rho^{3/2}}{(m_\rho^2-m_{\pi^0}^2)^{3/2}}-\sqrt{\Gamma(\omega\to\pi^0\gamma)}\frac{m_\omega^{3/2}}{(m_\omega^2-m_{\pi^0}^2)^{3/2}}\right)\,,\nonumber\\
 |A_\text{PC}^{\text{III},V^0}|&=\frac1{m_{D^0}+m_{V^0}}a_2\left(-f_\rho^2+\frac13f_\omega^2-\frac{V_{cs}^*V_{us}}{V_{cd}^*V_{ud}}\frac23f_\phi^2\right)V^{V^0}(0)\,,\nonumber\\
 |A_\text{PV}^{\text I,V^0}|&=0\,,\nonumber\\
 |A_\text{PV}^{\text{II},V^0}|&=-\frac1{\sqrt2(m_{D^0}^2-m_{V^0}^2)}a_2f_\rho\left(f_\rho A_1^\rho(m_{V^0}^2)(m_{D^0}+m_\rho)+f_\omega A_1^\omega(m_{V^0}^2)\frac{(m_{D^0}+m_\omega)m_\rho}{3m_\omega}\right)\,,\nonumber\\
 |A_\text{PV}^{\text{III},V^0}|&=\frac1{m_{D^0}-m_{V^0}}a_2\left(-f_\rho^2+\frac13f_\omega^2-\frac{V_{cs}^*V_{us}}{V_{cd}^*V_{ud}}\frac23f_\phi^2\right)A_1^{V^0}(0)
\end{align}
 and for $D^+ \to \rho^+ \gamma$
\begin{align}
 |A_\text{PC}^{\text I,\rho^+}|&=\frac{m_{{D^*}^+}^{7/2}\left(m_{{D^*}^+}^2-(m_{D^+}+m_{\pi^0})^2\right)^{3/4}\left(m_{{D^*}^+}^2-(m_{D^+}-m_{\pi^0})^2\right)^{3/4}m_\rho}{\sqrt{\pi\alpha_e}m_{D^+}^{3/2}(m_{{D^*}^+}^2-m_{D^+}^2)^{3/2}(m_{{D^*}^+}+m_{D^+}-m_{\pi^+})(m_{{D^*}^+}^2-m_\rho^2)}\nonumber\\
 &\times a_1f_\rho\sqrt{\frac{\Gamma({D^*}^+\to D^+\gamma)}{\Gamma({D^*}^+\to D^+\pi^0)}}f_+(0)\,,\nonumber\\
 |A_\text{PC}^{\text{II},\rho^+}|&=\frac{2\sqrt6m_{D^+}^2m_\rho^{3/2}}{\sqrt{\alpha_e}(m_{D^+}^2-m_{\pi^+}^2)(m_\rho^2-m_{\pi^+}^2)^{3/2}}a_1f_Df_\pi\sqrt{\Gamma(\rho^+\to\pi^+\gamma)}\,,\nonumber\\
 |A_\text{PC}^{\text{III},\rho^+}|&=\frac1{m_{D^+}+m_\rho}a_2\left(-f_\rho^2+\frac13f_\omega^2-\frac{V_{cs}^*V_{us}}{V_{cd}^*V_{ud}}\frac23f_\phi^2\right)V^\rho(0)\,,\nonumber\\
 |A_\text{PV}^{\text I,\rho^+}|&=\frac{2m_\rho}{m_{D^+}^2-m_\rho^2}a_1f_Df_\rho\,,\nonumber\\
 |A_\text{PV}^{\text{II},\rho^+}|&=\frac1{m_{D^+}^2-m_\rho^2}a_1f_\rho\left(f_\rho A_1^\rho(m_\rho^2)(m_{D^+}+m_\rho)-f_\omega A_1^\omega(m_\rho^2)\frac{(m_{D^+}+m_\omega)m_\rho}{3m_\omega}\right)\,,\nonumber\\
 |A_\text{PV}^{\text{III},\rho^+}|&=\frac1{m_{D^+}-m_\rho}a_2\left(-f_\rho^2+\frac13f_\omega^2-\frac{V_{cs}^*V_{us}}{V_{cd}^*V_{ud}}\frac23f_\phi^2\right)A_1^\rho(0)\,,
\end{align}
where $a_{1,2}$ are given in section \ref{sec:pheno} and $f_+(0)=(0.1426\pm0.0019)/|V_{cd}|$ \cite{Amhis:2016xyh}, where statistical and systematic uncertainties are added in quadrature.
For $D_s\to{K^*}^+\gamma$ decays
\begin{align}
 |A_\text{PC}^{\text I,{K^*}^+}|&=\frac{4m_{D_s^*}^{3/2}m_{{K^*}^+}}{m_{D_s}^{1/2}(m_{D_s^*}^2-m_{{K^*}^+}^2)}a_1f_{D_s}f_{K^*}\left|\lambda'+\lambda\tilde g_V\frac{-f_\phi}{3\sqrt2m_\phi}\right|\,,\nonumber\\
 |A_\text{PC}^{\text{II},{K^*}^+}|&=\frac{2\sqrt6m_{D_s}^2m_{{K^*}^+}^{3/2}}{\sqrt{\alpha_e}(m_{D_s}^2-m_{K^+}^2)(m_{{K^*}^+}^2-m_{K^+}^2)^{3/2}}a_1f_{D_s}f_K\sqrt{\Gamma({K^*}^+\to K^+\gamma)}\,,\nonumber\\
 |A_\text{PC}^{\text{III},{K^*}^+}|&=\frac1{m_{D_s}+m_{{K^*}^+}}a_2\left(-\frac{V_{cd}^*V_{ud}}{V_{cs}^*V_{us}}\left(-f_\rho^2+\frac13f_\omega^2\right)+\frac23f_\phi^2\right)V^{K^*}(0)\,,\nonumber\\
 |A_\text{PV}^{\text I,{K^*}^+}|&=\frac{2m_{{K^*}^+}}{m_{D_s}^2-m_{{K^*}^+}^2}a_1f_{D_s}f_{K^*}\,,\nonumber\\
 |A_\text{PV}^{\text{II},{K^*}^+}|&=\frac{2(m_{D_s}+m_\phi)m_{{K^*}^+}}{3m_\phi(m_{D_s}^2-m_{{K^*}^+}^2)}a_1f_{K^*}f_\phi A_1^{(D_s\to\phi)}(m_{{K^*}^+}^2)\,,\nonumber\\
 |A_\text{PV}^{\text{III},{K^*}^+}|&=\frac1{m_{D_s}-m_{{K^*}^+}}a_2\left(-\frac{V_{cd}^*V_{ud}}{V_{cs}^*V_{us}}\left(-f_\rho^2+\frac13f_\omega^2\right)+\frac23f_\phi^2\right)A_1^{K^*}(0)\,.
\end{align}
As the approximation of \cite{Fajfer:1997bh,Fajfer:1998dv} is not applicable for $\mathcal B(D_s^*\to D_s\pi^0)$ as a normalization mode due to isospin breaking \cite{Cho:1994zu} and $\Gamma_{D_s^*}$ is not measured, $\lambda'$ and $\lambda\tilde g_V$ in $A_\text{PC}^{\text I,{K^*}^+}$ are related to
\begin{align}
  \left|\lambda'+\lambda\tilde g_V\left(\pm\frac{f_\rho}{2\sqrt2m_\rho}+\frac{f_\omega}{6\sqrt2m_\omega}\right)\right|&=\frac{m_{D^*}^2\left(m_{D^*}^2-(m_D+m_{\pi^0})^2\right)^{3/4}\left(m_{D^*}^2-(m_D-m_{\pi^0})^2\right)^{3/4}}{4\sqrt{\pi\alpha_e}m_D(m_{D^*}^2-m_D^2)^{3/2}(m_{D^*}+m_D-m_{\pi^{0/+}})}\nonumber\\
  &\times\frac{f_+(0)}{f_D}\sqrt{\frac{\Gamma({D^*}^{0/+}\to D^{0/+}\gamma)}{\Gamma({D^*}^{0/+}\to D^{0/+}\pi^0)}}
\end{align}
and the ambiguity is fixed by means of the form factor $V$ \cite{Casalbuoni:1992dx}
\begin{align}
 \lambda\tilde g_V=\frac{V^V(0)}{f_D}\frac{\sqrt2m_{D^*}^2}{(m_D+m_V)(m_{D^*}+m_D-m_V)}\,.
\end{align}

Furthermore,
\begin{align}
 |A_\text{PC}^{\text I,D^0\to\phi\gamma}|&=-\frac{m_{{D^*}^0}^{7/2}\left(m_{{D^*}^0}^2-(m_{D^0}+m_{\pi^0})^2\right)^{3/4}\left(m_{{D^*}^0}^2-(m_{D^0}-m_{\pi^0})^2\right)^{3/4}m_\phi}{\sqrt{\pi\alpha_e}m_{D^0}^{3/2}(m_{{D^*}^0}^2-m_{D^0}^2)^{3/2}(m_{{D^*}^0}+m_{D^0}-m_{\pi^0})(m_{{D^*}^0}^2-m_\phi^2)}\nonumber\\
 &\times a_2f_\phi\sqrt{\frac{\Gamma({D^*}^0\to D^0\gamma)}{\Gamma({D^*}^0\to D^0\pi^0)}}f_+(0)\,,\nonumber\\
 |A_\text{PC}^{\text{II},D^0\to\phi\gamma}|&=-\frac{8m_{D^0}^2}{3(m_{D^0}^2-m_{K^0}^2)m_\phi}a_2f_Df_\phi|C_{VV\Pi}|\,,\nonumber\\
 |A_\text{PV}^{\text I,D^0\to\phi\gamma}|&=0\,,\nonumber\\
 |A_\text{PV}^{\text{II},D^0\to\phi\gamma}|&=-\frac{m_\phi}{m_{D^0}^2-m_\phi^2}a_2f_\phi\left(f_\rho A_1^\rho(m_\phi^2)\frac{m_{D^0}+m_\rho}{m_\rho}+f_\omega A_1^\omega(m_\phi^2)\frac{m_{D^0}+m_\omega}{3m_\omega}\right)\,,
\end{align}
where
\begin{align}
 |C_{VV\Pi}|=\frac{\sqrt6m_{{K^*}^{0/+}}^{3/2}}{\sqrt{\alpha_e}(m_{{K^*}^{0/+}}^2-m_{K^{0/+}}^2)^{3/2}}\frac{f_K}{\left|\mp\frac{f_\rho}{m_\rho}+\frac{f_\omega}{3m_\omega}-\frac{2f_\phi}{3m_\phi}\right|}\sqrt{\Gamma({K^*}^{0/+}\to K^{0/+}\gamma)}
\end{align}
yields $|C_{VV\Pi}|\simeq0.3$,
\begin{align}
 |A_\text{PC}^{\text I,D^0\to\bar K^{*0}\gamma}|&=-\frac{m_{{D^*}^0}^{7/2}\left(m_{{D^*}^0}^2-(m_{D^0}+m_{\pi^0})^2\right)^{3/4}\left(m_{{D^*}^0}^2-(m_{D^0}-m_{\pi^0})^2\right)^{3/4}m_{{K^*}^0}}{\sqrt{\pi\alpha_e}m_{D^0}^{3/2}(m_{{D^*}^0}^2-m_{D^0}^2)^{3/2}(m_{{D^*}^0}+m_{D^0}-m_{\pi^0})(m_{{D^*}^0}^2-m_{{K^*}^0}^2)}\nonumber\\
 &\times a_2f_{K^*}\sqrt{\frac{\Gamma({D^*}^0\to D^0\gamma)}{\Gamma({D^*}^0\to D^0\pi^0)}}f_+(0)\,,\nonumber\\
 |A_\text{PC}^{\text{II},D^0\to\bar K^{*0}\gamma}|&=-\frac{2\sqrt6m_{D^0}^2m_{{K^*}^0}^{3/2}}{\sqrt{\alpha_e}(m_{D^0}^2-m_{K^0}^2)(m_{{K^*}^0}^2-m_{K^0}^2)^{3/2}}a_2f_Df_K\sqrt{\Gamma({K^*}^0\to K^0\gamma)}\,,\nonumber\\
 |A_\text{PV}^{\text I,D^0\to\bar K^{*0}\gamma}|&=0\,,\\
 |A_\text{PV}^{\text{II},D^0\to\bar K^{*0}\gamma}|&=-\frac{m_{{K^*}^0}}{m_{D^0}^2-m_{{K^*}^0}^2}a_2f_{K^*}\left(f_\rho A_1^\rho(m_{{K^*}^0}^2)\frac{m_{D^0}+m_\rho}{m_\rho}+f_\omega A_1^\omega(m_{{K^*}^0}^2)\frac{m_{D^0}+m_\omega}{3m_\omega}\right)\,,  \nonumber
\end{align}
$|A_\text{PC,PV}^{\text{I,II},D^0\to{K^*}^0\gamma}|=|V_{cd}^*V_{us}/(V_{cs}^*V_{ud})|\,|A_\text{PC,PV}^{\text{I,II},D^0\to\bar K^{*0}\gamma}|$,
\begin{align}
 |A_\text{PC}^{\text I,D^+\to{K^*}^+\gamma}|&=\frac{m_{{D^*}^+}^{7/2}\left(m_{{D^*}^+}^2-(m_{D^+}+m_{\pi^0})^2\right)^{3/4}\left(m_{{D^*}^+}^2-(m_{D^+}-m_{\pi^0})^2\right)^{3/4}m_{{K^*}^+}}{\sqrt{\pi\alpha_e} m_{D^+}^{3/2}(m_{{D^*}^+}^2-m_{D^+}^2)^{3/2}(m_{{D^*}^+}+m_{D^+}-m_{\pi^+})(m_{{D^*}^+}^2-m_{{K^*}^+}^2)}\nonumber\\
 &\times a_1f_{K^*}\sqrt{\frac{\Gamma({D^*}^+\to D^+\gamma)}{\Gamma({D^*}^+\to D^+\pi^0)}}f_+(0)\,,\nonumber\\
 |A_\text{PC}^{\text{II},D^+\to{K^*}^+\gamma}|&=\frac{2\sqrt6m_{D^+}^2m_{{K^*}^+}^{3/2}}{\sqrt{\alpha_e}(m_{D^+}^2-m_{K^+}^2)(m_{{K^*}^+}^2-m_{K^+}^2)^{3/2}}a_1f_Df_K\sqrt{\Gamma(K^*\to K^+\gamma)}\,,\nonumber\\
 |A_\text{PV}^{\text I,D^+\to{K^*}^+\gamma}|&=\frac{2m_{{K^*}^+}}{m_{D^+}^2-m_{{K^*}^+}^2}a_1f_Df_{K^*}\,,\\
 |A_\text{PV}^{\text{II},D^+\to{K^*}^+\gamma}|&=\frac{m_{{K^*}^+}}{m_{D^+}^2-m_{{K^*}^+}^2}a_1f_{K^*}\left(f_\rho A_1^\rho(m_{{K^*}^+}^2)\frac{m_{D^+}+m_\rho}{m_\rho}-f_\omega A_1^{(D_s\to K^*)}(m_{{K^*}^+}^2)\frac{m_{D^+}+m_\omega}{3m_\omega}\right)  \nonumber
\end{align}
and
\begin{align}
 |A_\text{PC}^{\text I,D_s\to\rho^+\gamma}|&=\frac{4m_{D_s^*}^{3/2}m_\rho}{m_{D_s}^{1/2}(m_{D_s^*}^2-m_\rho^2)}a_1f_{D_s}f_\rho\left|\lambda'+\lambda\tilde g_V\frac{-f_\phi}{3\sqrt2m_\phi}\right|\,,\nonumber\\
 |A_\text{PC}^{\text{II},D_s\to\rho^+\gamma}|&=\frac{2\sqrt6m_{D_s}^2m_\rho^{3/2}}{\sqrt{\alpha_e}(m_{D_s}^2-m_{\pi^+}^2)(m_\rho^2-m_{\pi^+}^2)^{3/2}}a_1f_{D_s}f_\pi\sqrt{\Gamma(\rho^+\to\pi^+\gamma)}\,,\nonumber\\
 |A_\text{PV}^{\text I,D_s\to\rho^+\gamma}|&=\frac{2m_\rho}{m_{D_s}^2-m_\rho^2}a_1f_{D_s}f_\rho\,,\nonumber\\
 |A_\text{PV}^{\text{II},D_s\to\rho^+\gamma}|&=\frac{2(m_{D_s}+m_\phi)m_\rho}{3m_\phi(m_{D_s}^2-m_\rho^2)}a_1f_\rho f_\phi A_1^{(D_s\to\phi)}(m_{{K^*}^+}^2)\,.
\end{align}

\section{$C_7^{\rm eff}$ from  \texorpdfstring{$\langle Q_{3-6}\rangle$}{Q36} at two-loop QCD}\label{app:C7eff}

We obtain contributions to $C_7^\text{eff}$ from  the two-loop QCD matrix elements of $Q_{3-6}$ from \cite{Buras:2002tp}
\begin{align}
 C_7^{\text{eff}}\big|_{\langle Q_{3-6}\rangle}(\mu_c)=(V_{cd}^*V_{ud}+V_{cs}^*V_{us})\frac{\alpha_s}{4\pi}\sum_{i=3}^6\left(r_i-\frac{\gamma_{i7}^{(1)\text{eff}}}2\ln\frac{\mu_c^2}{m_c^2}\right)C_i\,,
\end{align}
where
\begin{align}\label{gamma_1eff_7}
 \gamma_{37}^{(1)\text{eff}}=-\frac{128}{81}\,,\quad\gamma_{47}^{(1)\text{eff}}=\frac{592}{243}\,,\quad\gamma_{57}^{(1)\text{eff}}=\frac{12928}{81}\,,\quad\gamma_{67}^{(1)\text{eff}}=\frac{40288}{243}   \, ,
\end{align}
\begin{align}
 &r_3=-\frac{4784}{243}-\frac{16\pi}{3\sqrt3}-\frac{64}9X_b+2a(1)-4b(1)-\frac{112}{81}i\pi\,,\nonumber\\
 &r_4=\frac{1270}{729}+\frac{8\pi}{9\sqrt3}+\frac{32}{27}X_b-\frac13a(1)-\frac{10}3b(1)-4b(m_s^2/m_c^2)+\frac{248}{243}i\pi\,,\nonumber\\
 &r_5=-\frac{113360}{243}-\frac{64\pi}{3\sqrt3}-\frac{256}{9}X_b+32a(1)-64b(1)-\frac{1792}{81}i\pi\,,\nonumber\\
 &r_6=-\frac{60980}{729}+\frac{32\pi}{9\sqrt3}+\frac{128}{27}X_b+\frac{20}3a(1)-\frac{88}3b(1)+6a(m_s^2/m_c^2)-40b(m_s^2/m_c^2)+\frac{1520}{243}i\pi\,,
\end{align}
where we use $m_{u,d}=0$,
\begin{align}
 a(\rho)&=\frac{16}9\bigg(\left(\frac52-\frac13\pi^2-3\zeta_3+\left(\frac52-\frac34\pi^2\right)\ln\rho+\frac14\ln^2\rho+\frac1{12}\ln^3\rho\right)\rho\nonumber\\
 &+\left(\frac74+\frac23\pi^2-\frac12\pi^2\ln\rho-\frac14\ln^2\rho+\frac1{12}\ln^3\rho\right)\rho^2+\left(-\frac76-\frac14\pi^2+2\ln\rho-\frac34\ln^2\rho\right)\rho^3\nonumber\\
 &+\left(\frac{457}{216}-\frac5{18}\pi^2-\frac1{72}\ln\rho-\frac56\ln^2\rho\right)\rho^4+\left(\frac{35101}{8640}-\frac{35}{72}\pi^2-\frac{185}{144}\ln\rho-\frac{35}{24}\ln^2\rho\right)\rho^5\nonumber\\
 &+\left(\frac{67801}{8000}-\frac{21}{20}\pi^2-\frac{3303}{800}\ln\rho-\frac{63}{20}\ln^2\rho\right)\rho^6\nonumber\\
 &+i\pi\left(\left(2-\frac16\pi^2+\frac12\ln\rho+\frac12\ln^2\rho\right)\rho+\left(\frac12-\frac16\pi^2-\ln\rho+\frac12\ln^2\rho\right)\rho^2+\rho^3+\frac59\rho^4+\frac{49}{72}\rho^5+\frac{231}{200}\rho^6\right)\bigg)\nonumber\\
 &+\mathcal O\left(\rho^7\ln^2\rho\right)\,,\nonumber\\
 b(\rho)&=-\frac89\bigg(\left(-3+\frac16\pi^2-\ln\rho\right)\rho-\frac23\pi^2\rho^{3/2}+\left(\frac12+\pi^2-2\ln\rho-\frac12\ln^2\rho\right)\rho^2\nonumber\\
 &+\left(-\frac{25}{12}-\frac19\pi^2-\frac{19}{18}\ln\rho+2\ln^2\rho\right)\rho^3+\left(-\frac{1376}{225}+\frac{137}{30}\ln\rho+2\ln^2\rho+\frac23\pi^2\right)\rho^4\nonumber\\
 &+\left(-\frac{131317}{11760}+\frac{887}{84}\ln\rho+5\ln^2\rho+\frac53\pi^2\right)\rho^5+\left(-\frac{2807617}{97200}+\frac{16597}{540}\ln\rho+14\ln^2\rho+\frac{14}3\pi^2\right)\rho^6\nonumber\\
 &+i\pi\left(-\rho+(1-2\ln\rho)\rho^2+\left(-\frac{10}9+\frac43\ln\rho\right)\rho^3+\rho^4+\frac23\rho^5+\frac79\rho^6\right)\bigg)\nonumber\\
 &+\mathcal O\left(\rho^7\ln^2\rho\right)
\end{align}
and
\begin{align}
 a(1)\simeq4.0859+\frac49i\pi\,,\qquad b(1)\simeq0.0316+\frac4{81}i\pi\,,\qquad X_b\simeq-0.1684\,.
\end{align}

\end{document}